\documentclass[aps,prb,twocolumn,amsmath,amssymb,floatfix,eqsecnum,showpacs,nofootinbib]{revtex4}

\usepackage{graphicx}
\usepackage{subfigure}
\usepackage{latexsym}
\usepackage{amsmath,amssymb}
\usepackage{bm} 
\usepackage{color}
\usepackage{stackrel}
\usepackage{epsfig}

\definecolor{myblue}{rgb}{.93, .93, 1}
\newcommand*\mybluebox[1]{%
\colorbox{myblue}{\hspace{1em}#1\hspace{1em}}}
\newcommand*\myFbluebox[1]{%
\fcolorbox{black}{myblue}{\hspace{1em}#1\hspace{1em}}}

\newcommand{\bsub}{\begin{subequations}}
\newcommand{\esub}{\end{subequations}}
\newcommand{\beq}{\begin{empheq}[box=\mybluebox]{align}}
\newcommand{\beqF}{\begin{empheq}[box=\myFbluebox]{align}}

\newcommand{\vex}[1]{\bm{\mathrm{#1}}}

\newcommand{\Nabla}{\bm{\nabla}}

\newcommand{\pup}[1]{{\scriptscriptstyle{({#1})}}}

\begin{document}

\title{
	Chalker scaling, level repulsion, and conformal invariance in critically delocalized quantum matter:
	Disordered topological superconductors and artificial graphene}
\author{Yang-Zhi Chou}\email{yc26@rice.edu}
\author{Matthew S. Foster}
\affiliation{Department of Physics and Astronomy, Rice University, Houston, Texas 77005, USA}
\begin{abstract}
	We numerically investigate critically delocalized wavefunctions 
	in models of 2D Dirac fermions, subject to vector potential disorder. 
	These describe the surface states of 3D topological superconductors, 
	and can also be realized through long-range correlated bond randomness
	in artificial materials like molecular graphene. 
	A ``frozen'' regime can occur for strong disorder in these systems, wherein 
	a single wavefunction 
	presents		
	a few localized peaks separated by macroscopic 
	distances. Despite this rarefied spatial structure, 
	we find robust 
	correlations	
	between eigenstates at different energies,
	at both weak and strong disorder. 
	The associated level statistics are 
	always 
	approximately	
	Wigner-Dyson.
	The system 
	shows	
	generalized Chalker (quantum critical) scaling,
	even when individual states are quasilocalized in space. 
	We confirm analytical predictions for the density of states and multifractal spectra. For a single 
	Dirac valley, we establish that finite energy states show
	universal multifractal spectra
	consistent with the integer quantum Hall plateau transition. A single Dirac fermion at finite energy 
	can therefore behave as a 
	``Quantum Hall critical metal.''	 
	For the case of two valleys and non-abelian
	disorder, we verify predictions of conformal field theory.
	Our results for the non-abelian case imply that both delocalization and conformal invariance are
	topologically-protected for multivalley topological superconductor surface states. 
\end{abstract}
\maketitle

\section{Introduction \label{Sec: Intro}}

Strong disorder can localize all wavefunctions in a band of energies.\cite{Anderson58}
In a localized phase, states close in energy are peaked at spatially distant centers, implying vanishingly small overlap
of the corresponding probability densities. 
The associated statistics of nearest-neighbor level spacings is Poissonian, i.e.\ there is no level repulsion. 
These 
features imply the similarity of an Anderson insulator to an integrable dynamical system,\cite{HuseOganesyan13}
an idea ignited by studies of many-body localization.\cite{Basko06,Oganesyan07,Pal10}
By contrast, the 
states of a diffusive metal 
are associated with 
quantum ergodicity, exhibiting 
Wigner-Dyson level statistics.\cite{Sivan87}
Near a mobility edge, extended states show
quantum critical (Chalker) scaling\cite{Chalker88,Chalker1990253} 
in the overlap of wavefunction probabilities at different energies.\cite{Fyodorov97,Cuevas07}

In this paper, we examine critically delocalized states in the presence of weak and strong disorder. 
Such states arise under special circumstances in low dimensions, when protected by 
symmetries and/or 
topology.\cite{Huckestein95,Evers08_RMP} 
These states can display non-ergodic 
characteristics, including a ``frozen regime'' wherein
a single wavefunction can appear almost localized, exhibiting a \emph{few} isolated peaks separated
by macroscopic distances.\cite{Chamon96,Ryu01A,Carpentier01} In this case, since
individual 
wavefunctions
show a mixture of localized and critical features, one might expect a breakdown
of correlations between different states with nearby energies.
If it were to exist, such a phase could be termed a ``non-ergodic'' or 
glassy metal, and would signify a failure of the scaling theory of localization.
Possible realizations include the Bethe lattice,\cite{Biroli12,DeLuca14} the region above the 
many-body localization transition,\cite{Luca13} or the critical region of the 
Anderson-Mott metal-insulator transition.\cite{DobroBook}

The systems we study consist of 2D massless Dirac fermions coupled to random vector potential disorder.
These arise as the surface states of 3D topological superconductors,\cite{Schnyder08,Foster12} in the presence
of any surface disorder that respects time-reversal symmetry. (The vector potentials do not encode physical
magnetic fields, but instead couple to spin and/or valley currents of 
gapless
surface 
quasiparticles. These currents are time-reversal even). We consider models in classes AIII and CI, which 
respectively retain U$(1)$ and SU$(2)$ spin symmetry in every realization of disorder. 
Class AIII can also be a chiral topological insulator.\cite{Hosur10}

Specifically, we study a single valley Dirac fermion perturbed by an abelian vector potential,\cite{Ludwig94} 
which is the minimal surface state of an
AIII topological superconductor.\cite{Schnyder08}
It also arises as the low-energy description of a 2D tight-binding model
with long-range correlated random hopping,\cite{Motrunich02} a system that might
be realizable in molecular graphene.\cite{Gomes12}  

We numerically evaluate level spacing statistics, the global density of states (DoS),
and multifractal spectra\cite{Huckestein95,Evers08_RMP} of single particle wavefunctions. 
We also compute two-wavefunction correlations between states at different energies. 
Our work extends previous numerics\cite{Hatsugai97,Morita97,Ryu01B}
to stronger disorder (beyond the freezing transition). 
Prior numerical work on the strong disorder regime investigated the DoS\cite{Motrunich02}
and the multifractal spectrum of the exact zero energy wavefunction.\cite{Chen12}
Our work adds Chalker scaling, level statistics, and multifractal spectra
of the low-energy states.
Finally, we also investigate a model with two valleys in classes CI and AIII as the simplest 
example of a Dirac fermion subject to a non-abelian disorder potential.\cite{Nersesyan94}
In our finite size studies, we do not attempt to prove delocalization.
Instead, we match our results for the critical behavior of the DoS and multifractal spectra
to 
predictions for the critically delocalized states expected to form in these
systems.

\begin{figure}
\centering
\subfigure[$\Delta_A=0.4\pi$, weak disorder region]{
\includegraphics[width=0.4\textwidth]{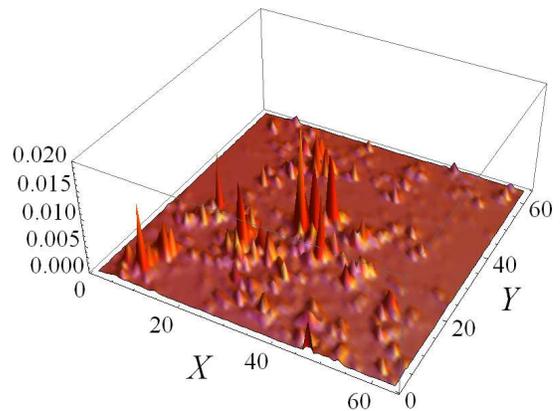}
	}\\[3mm]
\subfigure[$\Delta_A=3\pi$, ``frozen'' region]{
\includegraphics[width=0.4\textwidth]{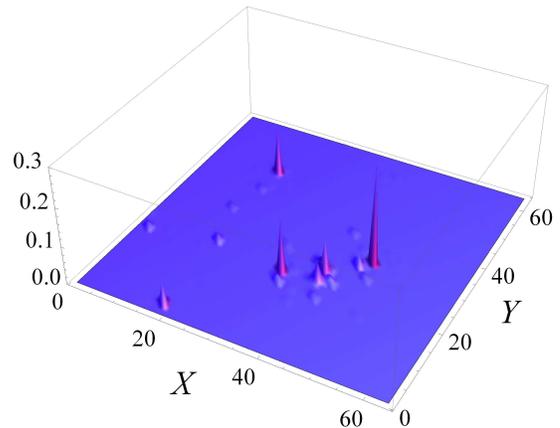}
	}
\caption{
Probability density of the exact zero energy wavefunctions in the single valley Dirac fermion model.\cite{Ludwig94} 
(a) represents the critically delocalized states in the weak disorder region ($\Delta_A<2\pi$). The distribution 
is spatially inhomogeneous and multifractal. (b) shows the spatial signature of the ``frozen'' states in the 
strong disorder region ($\Delta_A>2\pi$). The wavefunction is characterized by rarefied peaks. 
For both cases, we generate the analytical wavefunction\cite{Ludwig94} with a 64-by-64 spatial resolution.
\label{Zero_Wavefcn}}
\end{figure}

Many of the properties of the single-valley model are known analytically. 
The global DoS is critical. The corresponding dynamical exponent
is non-universal and depends on the strength of disorder.\cite{Ludwig94}
The multifractal spectra of the low-energy wavefunctions can be obtained 
exactly.\cite{Ludwig94} There is a ``freezing transition'' driven by the disorder 
strength in the low-energy states,\cite{Chamon96,Castillo97,Carpentier01}
beyond which individual wavefunctions become quasilocalized. The 
low energy global DoS is also modified in this regime.\cite{Motrunich02,Horovitz02,Mudry03}
Despite this, at the Dirac point the dc (zero temperature, Landauer) conductance is 
a universal number $e^2/\pi h$, valid for arbitrary disorder strength.\cite{Ludwig94}
See Fig.~\ref{Zero_Wavefcn} for a comparison of wavefunctions at weak and strong disorder.

Our results imply that 
energetic correlations survive in this system, even for strong disorder. In particular, 
after taking into account the critical behavior of the global DoS, we show that the 
level spacing statistics remain 
approximately	
Wigner-Dyson, below\cite{Morita97} and above the freezing 
transition. 
Using a long-range correlated random hopping model to simulate the 
low-energy Dirac fermion physics,\cite{Motrunich02} we confirm that the overlap between wavefunction 
probabilities at different energies exhibits a generalized form of Chalker 
scaling.\cite{Chalker88,Chalker1990253,Huckestein94,Pracz96,Fyodorov97,Cuevas07,Kravtsov10} 
This also holds below and above freezing, and	
implies that while individual states become highly rarified in space in the frozen regime,
these 
remain strongly correlated in energy. We conclude that a non-ergodic metal 
as defined above is not realized here. 
Strong correlations between nearby eigenstates 
with rarified structure were also
demonstrated in the sparse random matrix model.\cite{Fyodorov97}
We conjecture that ``non-ergodic'' signatures in energy 
(Poissonian level statistics, breakdown of Chalker scaling) for single particle states 
can occur only inside a true Anderson insulator.

We also show that strong disorder has a much weaker and universal effect at 
larger energies, wherein the multifractal statistics cross over to those of the 
integer quantum Hall plateau transition, consistent with previous 
work.\cite{Ludwig94,Ostrovsky07,Nomura07_QHE}
For the non-abelian two-valley model, we confirm predictions of 
conformal field theory.\cite{Nersesyan94,Mudry96,Caux96}
Our results for the non-abelian case imply that both 
delocalization and conformal invariance are topologically-protected 
for multivalley topological superconductor surface states.
At the surface of a topological superconductor in class CI or AIII,
gapless quasiparticles are characterized by a well-defined spin conductance
(because spin is conserved). Strict conformal invariance is consistent 
with the universality of the Landauer spin conductance,\cite{Ostrovsky06}
as in the single valley case.\cite{Ludwig94}
The robustness of this result to interaction effects will be explored
elsewhere.\cite{Xie14}
 
The rest of the article 
is
organized as follows: The model of the single valley Dirac fermion and the 
numerical methods are introduced in 
Sec.~\ref{Sec: SymModels}.
We show agreement between the numerical results and the analytical predictions 
for the global DoS and the zero-energy multifractal spectra in 
Secs.~\ref{Sec: DynExp} and \ref{Sec: MFC},
respectively. 
Level spacing statistics and the correlations 
between wavefunctions at different energies are studied in 
Sec.~\ref{Sec: DynProp}.
The finite energy states of the single-valley model 
are
discussed in 
Sec.~\ref{Sec: QHPM}.
In 
Sec.~\ref{Sec: NA},
we investigate the two-valley model and confirm the conformal field theory 
predictions. 
We conclude with a discussion in Sec.~\ref{Sec: Disc}.

%%%%%%%%%%%%%%%%%%%%%%%%%%%%%%%%%%%%%%%%%%%%%%%%%%%%%%%%%%%%%%%%%%%%%%%%%%%%%%%%%%%%%%%%%
%%%%%%%%%%%%%%%%%%%%%%%%%%%%%%%%%%%%%%%%%%%%%%%%%%%%%%%%%%%%%%%%%%%%%%%%%%%%%%%%%%%%%%%%%
%%%%%%%%%%%%%%%%%%%%%%%%%%%%%%%%%%%%%%%%%%%%%%%%%%%%%%%%%%%%%%%%%%%%%%%%%%%%%%%%%%%%%%%%%
%%%%%%%%%%%%%%%%%%%%%%%%%%%%%%%%%%%%%%%%%%%%%%%%%%%%%%%%%%%%%%%%%%%%%%%%%%%%%%%%%%%%%%%%%
%%%%%%%%%%%%%%%%%%%%%%%%%%%%%%%%%%%%%%%%%%%%%%%%%%%%%%%%%%%%%%%%%%%%%%%%%%%%%%%%%%%%%%%%%
%%%%%%%%%%%%%%%%%%%%%%%%%%%%%%%%%%%%%%%%%%%%%%%%%%%%%%%%%%%%%%%%%%%%%%%%%%%%%%%%%%%%%%%%%
%%%%%%%%%%%%%%%%%%%%%%%%%%%%%%%%%%%%%%%%%%%%%%%%%%%%%%%%%%%%%%%%%%%%%%%%%%%%%%%%%%%%%%%%%
%%%%%%%%%%%%%%%%%%%%%%%%%%%%%%%%%%%%%%%%%%%%%%%%%%%%%%%%%%%%%%%%%%%%%%%%%%%%%%%%%%%%%%%%%
%%%%%%%%%%%%%%%%%%%%%%%%%%%%%%%%%%%%%%%%%%%%%%%%%%%%%%%%%%%%%%%%%%%%%%%%%%%%%%%%%%%%%%%%%
%%%%%%%%%%%%%%%%%%%%%%%%%%%%%%%%%%%%%%%%%%%%%%%%%%%%%%%%%%%%%%%%%%%%%%%%%%%%%%%%%%%%%%%%%
%%%%%%%%%%%%%%%%%%%%%%%%%%%%%%%%%%%%%%%%%%%%%%%%%%%%%%%%%%%%%%%%%%%%%%%%%%%%%%%%%%%%%%%%%
%%%%%%%%%%%%%%%%%%%%%%%%%%%%%%%%%%%%%%%%%%%%%%%%%%%%%%%%%%%%%%%%%%%%%%%%%%%%%%%%%%%%%%%%%
%%%%%%%%%%%%%%%%%%%%%%%%%%%%%%%%%%%%%%%%%%%%%%%%%%%%%%%%%%%%%%%%%%%%%%%%%%%%%%%%%%%%%%%%%
%%%%%%%%%%%%%%%%%%%%%%%%%%%%%%%%%%%%%%%%%%%%%%%%%%%%%%%%%%%%%%%%%%%%%%%%%%%%%%%%%%%%%%%%%
%%%%%%%%%%%%%%%%%%%%%%%%%%%%%%%%%%%%%%%%%%%%%%%%%%%%%%%%%%%%%%%%%%%%%%%%%%%%%%%%%%%%%%%%%
%%%%%%%%%%%%%%%%%%%%%%%%%%%%%%%%%%%%%%%%%%%%%%%%%%%%%%%%%%%%%%%%%%%%%%%%%%%%%%%%%%%%%%%%%

\section{Symmetry Properties and Models \label{Sec: SymModels}}

Dirac fermions in solid state systems can emerge from graphene(-like) materials\cite{Neto09_RMP,Das_Sarma11_RMP} 
and the surfaces of 3D topological matter.\cite{Schnyder08,Hasan10_RMP,Qi11_RMP} In this section, we focus on 
a single valley Dirac fermion in 2D, subject to a random vector potential. This 
describes the surface of a 3D time-reversal symmetric topological superconductor with spin U$(1)$ symmetry
(class AIII),\cite{Schnyder08} with surface imperfections due to impurity atoms, vacancies, edge
and corner potentials, etc.
The Hamiltonian of the 2D Dirac fermion is 
\begin{align}
	\mathcal{H}
	&=\int d^2\vex{x}\,\psi^{\dagger}(\vex{x})\left[-i\bm{\sigma}\cdot\Nabla+\bm{\sigma}\cdot\vex{A}(\vex{x})\right]\psi(\vex{x})\label{H_RVPD}\\
	\nonumber&\equiv\int d^2\vex{x}\,\psi^{\dagger}(\vex{x})\hat{h}(\vex{x})\psi(\vex{x})
\end{align}
where $\vex A=({A}_x,{A}_y)$ is the vector potential and the Dirac pseudospin 
$\bm\sigma=(\sigma_x,\sigma_y)$. $\sigma_x$ and $\sigma_y$ 
are two of the three standard Pauli matrices. 

The Dirac Hamiltonian satisfies a chiral 
symmetry
condition
\begin{align}
	\sigma_z\hat{h}\sigma_z=-\hat{h}.	
	\label{chC}
\end{align}
Imposing chiral symmetry in every disorder realization
implies that the Hamiltonian only allows terms that couple to $\sigma_x$ and $\sigma_y$.

As mentioned above, 
Eq.~(\ref{H_RVPD}) can be 
viewed
as the surface state of a topological superconductor in class AIII. 
This is a superconductor with a remnant U(1) component of spin SU(2)
symmetry, as could arise due to bulk p-wave spin-triplet pairing.\cite{Foster08,Schnyder08}
The $z$ component of the physical spin is conserved and plays 
the	
role of U(1) 
charge in this representation.
In this intepretation, the pseudospin Pauli matrices 
$\{\sigma_{\mu}\}$
in Eq.~(\ref{H_RVPD}) act on a
combination of particle-hole and orbital degrees of freedom, 
but not on the physical spins.\cite{Foster14} 
Time-reversal and 
particle-hole symmetries 
combine to form the
chiral condition in 
Eq. (\ref{chC}). 
Any disorder terms obeying 
time reversal symmetry
will only appear in the form of vector potential $\vex A$
(up to irrelevant perturbations).	
Without loss of the generality, 
one typically considers 
zero-mean, white-noise-correlated potentials,
\bsub
\begin{align}
	\langle{A}_{\bar{\alpha}}(\vex{y})\rangle_{\text{dis}}&=0,\label{ZM}\\
	\langle{A}_{\bar{\alpha}}(\vex{y}){A}_{\bar{\alpha}'}(\vex{y}')\rangle_{\text{dis}}
	&=
	\Delta_{A} \, \delta_{\bar{\alpha},\bar{\alpha}'}\delta^{(2)}(\vex{y}-\vex{y}'),\label{WN}
\end{align}
\esub
where $\langle\dots\rangle_{\text{dis}}$ denotes disorder average, and $\Delta_{A}$ determines the disorder strength. 
In these equations,
$\bar{\alpha}$ and $\bar{\alpha}'$ 
span the $x$ and $y$ components. 

As discussed in Sec.~\ref{Sec: Intro}, many properties of this model are known analytically.
The dc conductance is universal,\cite{Ludwig94} but various physical quantities like the 
dynamical critical exponent
and the multifractal spectra 
of the low energy wavefunctions
depend on the 
strength of the disorder
$\Delta_A$.\cite{Ludwig94,Chamon96,Castillo97,Carpentier01}

\subsection{Momentum Space Formalism for Dirac fermions \label{Sec: MFD}}

In this section, we 
describe our 
numerical
momentum space formalism 
for Dirac fermions (MFD).
It is a direct way to simulate the single-valley model in the presence of random potentials.\cite{Nomura07_DF,Bardarson07} The energy cutoff $\Lambda$ is fixed in the MFD simulations.
The Fourier transform conventions are given by

\begin{align}
\nonumber\widetilde{\psi}_{\vex{n}}&
=\frac{1}{\sqrt{L^2}}\int d^2{\vex{x}}\,e^{-i\frac{2\pi}{L}\vex{n}\cdot\vex{x}}\psi(\vex{x}),\\
\nonumber\widetilde{{A}}_{\bar{\mu},\vex{n}}&
=\int d^2{\vex{x}}\,e^{-i\frac{2\pi}{L}\bm{n}\cdot{\vex{x}}}{A}_{\bar{\mu}}({\vex{x}}),
\end{align}
where $\vex{n}=(n_x,n_y)$ and $L$ is the length of the system size. 
We assume
periodic boundary conditions 
so that
$n_x$ and $n_y$ are integer-valued.

The Dirac Hamiltonian in the Fourier space is
\begin{align*}
	\mathcal{H}
	=&\,
	\frac{2\pi}{L}\sum_{\vex{n}}
	\widetilde{\psi}^{\dagger}_{\vex{n}}
	\left(\vex{n}\cdot\vex{\sigma}\right)
	\widetilde{\psi}_{\vex{n}}\\
	&\,
	+\frac{1}{L^2}\sum_{\vex{n},\vex{m}}
	\widetilde{\psi}^{\dagger}_{\vex{m}}
		\left[
		\widetilde{A}_{x,\vex{m}-\vex{n}} \, \sigma_x
		+\widetilde{A}_{y,\vex{m}-\vex{n}} \, \sigma_y
	\right]
	\widetilde{\psi}_{\vex{n}}
\end{align*}

In numerical simulations, we need to introduce two additional scales. 
These are the
cutoff in Fourier modes ($\mathcal{N}$), 
and the Gaussian correlation length of 
the
disorder potential ($\xi$). 
The Fourier modes $n_x$ and $n_y$ are constrained such that $-\mathcal{N}\le n_x,\,n_y\le\mathcal{N}$. 
The momentum grid 
has size
$(2\mathcal{N}+1)^2$. 
The total dimension of the Hilbert space is $2(2\mathcal{N}+1)^2$, where the extra factor of 2 
accounts for the Dirac pseudospin. 
We hold constant the energy cutoff $\Lambda=2\pi/r$, where
\begin{align}\label{rDef}
	r \equiv L/\mathcal{N}.
\end{align}
$r$ is about twice larger than the finest resolution 
in the calculations, $L/(2\mathcal{N}+1)$.

On the other hand, the white-noise correlation in Eq. (\ref{WN}) 
requires regularization. 
We replace the 
delta distribution 
with 
a random phase, fixed Gaussian amplitude distribution.
We parametrize the disorder potential via
\begin{align}\label{AParam}
	\widetilde{{A}}_{\bar{\mu},\vex{n}}
	=
	\sqrt{\Delta_A} \,
	L 
	\exp\left[{\textstyle{ - \frac{1}{4}\left(\frac{2\pi}{L}\vex{n}\xi\right)^2 }}\right]
	e^{i\theta_{\bar{\mu},\vex{n}}}.
\end{align}
where 
$\theta_{\bar{\mu},\vex{n}}\in[0,2\pi)$ 
is a random phase associated with 
$\widetilde{{A}}_{\bar{\mu},\vex{n}}$. We take 
$\theta_{\bar{\mu},\vex{n}}=-\theta_{\bar{\mu},-\vex{n}}$ 
because the $\vex{A}(\vex{x})$ is real-valued. 
The randomness is implemented by assigning a random phase to each Fourier mode. 
This approach is equivalent to the disorder average up to a finite size correction. 
We show the validity of the random phase method in the Appendix \ref{App: CorrDis}.
In Eq.~(\ref{AParam}), the correlation length $\xi$
is of the order of $r$ 
[Eq.~(\ref{rDef})]. 

\begin{figure}
\includegraphics[width=0.4\textwidth]{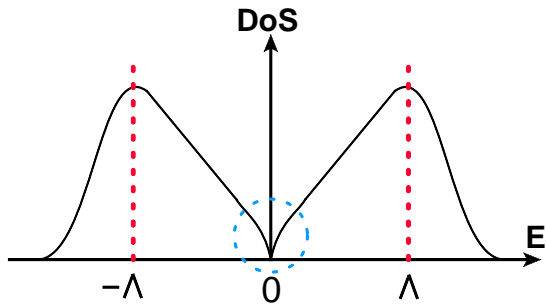}
\caption{Sketch of the DoS in the MFD, 
in the presence of disorder. 
The red dashed lines mark the position of the energy cutoff $\Lambda=\mathcal{N}(2\pi/L)$, which is fixed to a constant. 
The chiral 
region is circled by a blue dashed line,
wherein the dynamic critical exponent is modified by the disorder strength 
$\Delta_A$.\cite{Ludwig94,Motrunich02,Horovitz02,Mudry03} 
The states away from the chiral region but inside 
the cutoff 
typically show a DoS linear in energy, which is also the result for clean
2D Dirac fermions.
\label{DiracDoS}}
\end{figure}

In Fig. (\ref{DiracDoS}), 
we sketch
the DoS 
as a function of the energy $E$
in the MFD.
High energy
states outside the cutoff (red dashed lines) are artifacts of the simulations. 
There is a region of states (blue circled region) in the vicinity of $E=0$ reflecting 
the zero-energy quantum critical behavior of the single-valley model. We term this 
the
``chiral region.'' 
The states 
at intermediate energies above the chiral region and below the cutoff exhibit
the linear DoS expected for clean 2D Dirac fermions.

The MFD approach is 
rather memory intensive
because the matrices in momentum space are very dense. 
The calculations are therefore restricted to small momentum grid sizes.

%%%%%%%%%%%%%%%%%%%%%%%%%%%%%%%%%%%%%%%%%%%%%%%%%%%%%%%%%%%%%%%%%%%%%%%%%%%%%%%%%%%%%%%%%%%%%%%
%%%%%%%%%%%%%%%%%%%%%%%%%%%%%%%%%%%%%%%%%%%%%%%%%%%%%%%%%%%%%%%%%%%%%%%%%%%%%%%%%%%%%%%%%%%%%%%
%%%%%%%%%%%%%%%%%%%%%%%%%%%%%%%%%%%%%%%%%%%%%%%%%%%%%%%%%%%%%%%%%%%%%%%%%%%%%%%%%%%%%%%%%%%%%%%
%%%%%%%%%%%%%%%%%%%%%%%%%%%%%%%%%%%%%%%%%%%%%%%%%%%%%%%%%%%%%%%%%%%%%%%%%%%%%%%%%%%%%%%%%%%%%%%
%%%%%%%%%%%%%%%%%%%%%%%%%%%%%%%%%%%%%%%%%%%%%%%%%%%%%%%%%%%%%%%%%%%%%%%%%%%%%%%%%%%%%%%%%%%%%%%
%%%%%%%%%%%%%%%%%%%%%%%%%%%%%%%%%%%%%%%%%%%%%%%%%%%%%%%%%%%%%%%%%%%%%%%%%%%%%%%%%%%%%%%%%%%%%%%

\subsection{Lattice Model}

As an alternative approach,
we study a random hopping model of spinless fermions on a bipartite lattice. The Hamiltonian is
\begin{align*}
	\mathcal{H}
	=
	\sum_{\langle \vex{r}_A,\vex{r}_B \rangle}
	\left[t_{\vex{r}_A,\vex{r}_B}c^{\dagger}_A(\vex{r}_A)c_B(\vex{r}_B)+\text{h.c.}\right],
\end{align*}
where 
$c^{\dagger}_{A,B}$ ($c_{A,B}$) is the creation (annihilation) operator, 
$\vex r_{A}$ ($\vex r_{B}$)
specifies the position of 
a 
point 
in
sublattice 
$A$ ($B$), and 
$t_{\vex{r}_A,\vex{r}_B}$ is the hopping 
amplitude between $\vex{r}_A$ and $\vex{r}_B$.
The sum runs over nearest-neighbor pairs of sites.

Similar to the Dirac Hamiltonian we wish to study [Eq.~(\ref{H_RVPD})],
the hopping problem on bipartite lattices defined above satisfies a chiral symmetry 
(also called sublattice symmetry) at half filling.\cite{Gade1991213,Gade1993499} 
Moreover, 
Dirac fermions can emerge in the low-energy description for specific bipartite lattices
(i.e., the honeycomb lattice and the square lattice with $\pi$-flux).\cite{Hatsugai97,Motrunich02}

Unfortunately, sublattice symmetry and low-energy Dirac fermions are insufficient to realize
Eq.~(\ref{H_RVPD}). The latter describes the surface states of a bulk topological superconductor,
which one expects cannot be faithfully realized in a microscopic 2D system.\cite{Schnyder08,Hasan10_RMP,Qi11_RMP}
For example, the half-filled 
honeycomb lattice 
model
with bond randomness 
has an effective description in terms of
Dirac fermions with 
random vector and Kekul\'e\cite{Hou07} mass potentials. 
The low-energy theory is
\begin{align}
	\mathcal{H}
	\approx&\, v_F\int d^2{\vex{x}}\,\psi^{\dagger}\left[-i\sigma_x\partial_x-i\sigma_y\partial_y+\vex{A}\cdot\vex{\sigma}\kappa_z\right]\psi\label{lowE}
	\\
	\nonumber
	&\,
	+\int d^2{\vex{x}}\,\psi^{\dagger}\left[m_x\sigma_z\kappa_x+m_y\sigma_z\kappa_y\right]\psi,
\end{align}
where $v_F$ is the Fermi velocity, $\psi$ is the Dirac field [Eq. (\ref{honeycomb_psi}) in the Appendix 
\ref{App: Pi-flux}], and $\kappa_z$ is the valley Pauli matrix. 
If the system is translationally and rotationally invariant on average, then the mean value
of the vector and mass potentials vanish.
However, any non-zero variance of the
Kekul\'e mass terms ($m_x$ and $m_y$) 
drives the system into the Gade-Wegner fixed point.\cite{Gade1991213,Gade1993499} 
This is characterized by a
divergent DoS,
\begin{align}
	\nu(E)\sim\frac{1}{E}\exp(-c\left|\ln E\right|^{\alpha}),\label{DoS_Gade}
\end{align}
where $c$ is a non-universal constant. 
The exponent $\alpha$ takes the value $1/2$ at intermediate energies,\cite{Gade1991213,Gade1993499} and crosses over
to $2/3$ as $E \rightarrow 0$.\cite{Horovitz02,Motrunich02,Mudry03} 
This is different from the Dirac model in Eq.~(\ref{H_RVPD}), which exhibits a $\Delta_A$-dependent power law 
density of states.

\begin{figure}
\centering
\subfigure[Square lattice with $\pi$-flux]{
\includegraphics[width=0.2\textwidth]{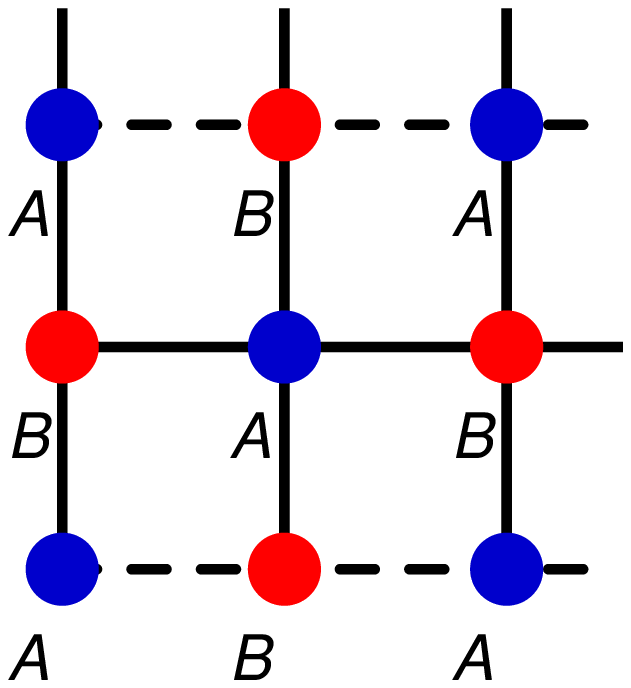}
	}
\subfigure[Honeycomb lattice]{
\includegraphics[width=0.2\textwidth]{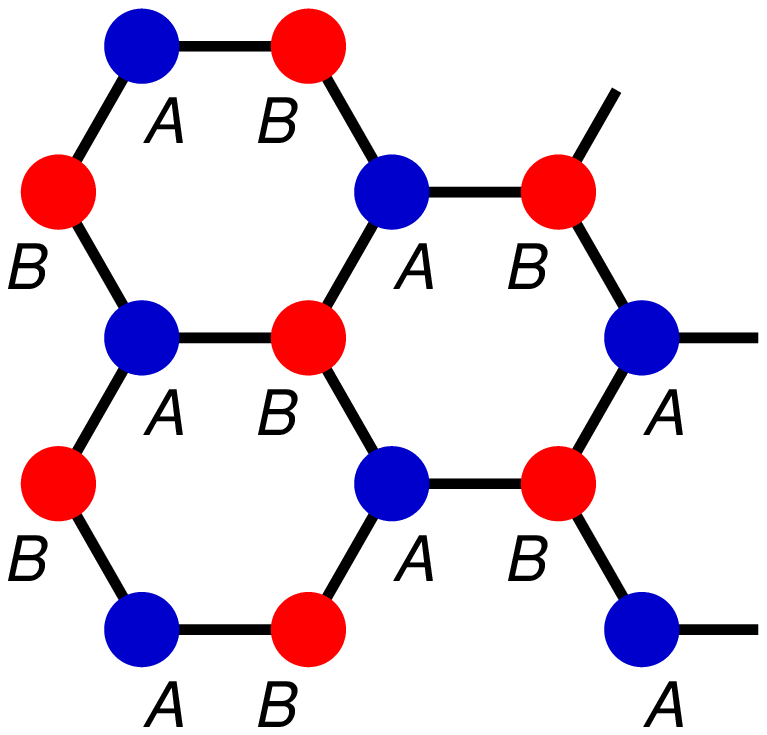}
	}
\caption{
Bipartite lattices with low-energy Dirac fermions:
square lattice with $\pi$-flux and 
the
honeycomb lattice. 
Labels
$A$ and $B$ indicate the sublattices. 
In the clean limit, the homogeneous hopping amplitudes $t_{\vex{r}_A,\vex{r}_B}^{\pup{0}}$ [Eq.~(\ref{t_MDH})] 
are 
equal to
$+1$ for solid lines and $-1$ for dashed lines.
\label{MDH_lattice}}
\end{figure}

A way to avoid 
Gade-Wegner physics is to implement the long-range correlated random hopping proposed 
by Motrunich, Damle, and Huse (MDH) 
in Ref.~\onlinecite{Motrunich02}. 
The 
MDH
construction is valid for any bipartite lattice with 
emergent
Dirac fermions. 
One defines a
real-valued logarithmic correlated potential $V(\vex{y})$ via
\begin{align}
	\langle V(\vex{y})V(\vex y')\rangle_{\text{dis}}
	=
	-\frac{\Delta_A}{2\pi}\ln\left(\frac{|\vex y-\vex y'|}{a}\right),
	\label{logCor}
\end{align}
where $a$ is some short distance scale. The hopping amplitudes are generated by
\begin{align}
	t_{\vex{r}_A,\vex{r}_B}=t_{\vex{r}_B,\vex{r}_A}
	=
	e^{V(\vex{r}_A)}
	t_{\vex{r}_A,\vex{r}_B}^\pup{0}	
	e^{-V(\vex{r}_B)},
	\label{t_MDH}
\end{align}
where $\vex{r}_A$ and $\vex{r}_B$ correspond to 
nearest-neighbor sites on the A and B sublattices, as depicted
in Fig.~\ref{MDH_lattice} for both the 
($\pi$-flux)
square 
and
honeycomb lattices.

The log-correlated disorder is
smooth 
on the lattice
scale.
Thus, the difference of disorder potentials at nearby positions can be approximated 
as
$
	V(\vex{y} + \vex{v}) - V(\vex{y}) \approx (\vex{v}\cdot\Nabla)V(\vex{y})
$. 
The low-energy theory can be derived by throwing away second and higher order derivative terms. 
Mass terms vanish in the naive long wavelength limit. One can show that 
${A}_x
\approx
\partial_y V$ 
and 
${A}_y
\approx
-\partial_x V$ 
in 
Eq. (\ref{lowE}). The random vector potential 
$\vex{A}$ generated this way satisfies 
Eqs.~(\ref{ZM}) and (\ref{WN}). 
The low energy theory of the MDH model describes two 
(nearly)
decoupled Dirac fermions with random vector potentials.

It is also important to discuss on the specific coarse graining conditions. 
On the honeycomb lattice, the
Kekul\'e masses 
correspond
to certain period-3 hopping patterns.\cite{Hou07} 
The
proper coarse graining cell should be at 
least 
as large as
a 
hexagonal plaquette
(6 sites, including two sublattices) 
on the honeycomb lattice. On the contrary, the minimum coarse graining cell on the square lattice with 
$\pi$-flux is a 2-by-2 block (see Appendix \ref{App: Pi-flux}). 
We mainly study 
the
MDH model on 
the
square lattice with $\pi$-flux for convenience.

%%%%%%%%%%%%%%%%%%%%%%%%%%%%%%%%%%%%%%%%%%%%%%%%%%%%%%%%%%%%%%%%%%%%%%%%%%%%%%%%%%%%%%%%%
%%%%%%%%%%%%%%%%%%%%%%%%%%%%%%%%%%%%%%%%%%%%%%%%%%%%%%%%%%%%%%%%%%%%%%%%%%%%%%%%%%%%%%%%%
%%%%%%%%%%%%%%%%%%%%%%%%%%%%%%%%%%%%%%%%%%%%%%%%%%%%%%%%%%%%%%%%%%%%%%%%%%%%%%%%%%%%%%%%%
%%%%%%%%%%%%%%%%%%%%%%%%%%%%%%%%%%%%%%%%%%%%%%%%%%%%%%%%%%%%%%%%%%%%%%%%%%%%%%%%%%%%%%%%%
%%%%%%%%%%%%%%%%%%%%%%%%%%%%%%%%%%%%%%%%%%%%%%%%%%%%%%%%%%%%%%%%%%%%%%%%%%%%%%%%%%%%%%%%%
%%%%%%%%%%%%%%%%%%%%%%%%%%%%%%%%%%%%%%%%%%%%%%%%%%%%%%%%%%%%%%%%%%%%%%%%%%%%%%%%%%%%%%%%%
%%%%%%%%%%%%%%%%%%%%%%%%%%%%%%%%%%%%%%%%%%%%%%%%%%%%%%%%%%%%%%%%%%%%%%%%%%%%%%%%%%%%%%%%%
%%%%%%%%%%%%%%%%%%%%%%%%%%%%%%%%%%%%%%%%%%%%%%%%%%%%%%%%%%%%%%%%%%%%%%%%%%%%%%%%%%%%%%%%%
%%%%%%%%%%%%%%%%%%%%%%%%%%%%%%%%%%%%%%%%%%%%%%%%%%%%%%%%%%%%%%%%%%%%%%%%%%%%%%%%%%%%%%%%%
%%%%%%%%%%%%%%%%%%%%%%%%%%%%%%%%%%%%%%%%%%%%%%%%%%%%%%%%%%%%%%%%%%%%%%%%%%%%%%%%%%%%%%%%%
%%%%%%%%%%%%%%%%%%%%%%%%%%%%%%%%%%%%%%%%%%%%%%%%%%%%%%%%%%%%%%%%%%%%%%%%%%%%%%%%%%%%%%%%%
%%%%%%%%%%%%%%%%%%%%%%%%%%%%%%%%%%%%%%%%%%%%%%%%%%%%%%%%%%%%%%%%%%%%%%%%%%%%%%%%%%%%%%%%%
%%%%%%%%%%%%%%%%%%%%%%%%%%%%%%%%%%%%%%%%%%%%%%%%%%%%%%%%%%%%%%%%%%%%%%%%%%%%%%%%%%%%%%%%%
%%%%%%%%%%%%%%%%%%%%%%%%%%%%%%%%%%%%%%%%%%%%%%%%%%%%%%%%%%%%%%%%%%%%%%%%%%%%%%%%%%%%%%%%%
%%%%%%%%%%%%%%%%%%%%%%%%%%%%%%%%%%%%%%%%%%%%%%%%%%%%%%%%%%%%%%%%%%%%%%%%%%%%%%%%%%%%%%%%%
%%%%%%%%%%%%%%%%%%%%%%%%%%%%%%%%%%%%%%%%%%%%%%%%%%%%%%%%%%%%%%%%%%%%%%%%%%%%%%%%%%%%%%%%%

\section{Dynamical Exponent and Density of States \label{Sec: DynExp}}

An important analytical result for the 
single-valley model is the exact disorder dependence\cite{Ludwig94,Horovitz02,Motrunich02,Mudry03} 
of the dynamic critical exponent $z$,
\begin{align}
	z
	=
	\begin{cases}{}
		1+\frac{\Delta_A}{\pi},& \Delta_A\le 2\pi\\[2mm]
		4\sqrt{\frac{\Delta_A}{2\pi}}-1, & \Delta_A> 2\pi
	\end{cases}
	\label{dyn_exp}
\end{align}
The dynamical exponent shows a non-analyticity at $\Delta_A=2\pi$, which signals 
a 
``freezing'' transition\cite{Chamon96,Horovitz02,Motrunich02,Mudry03}
for the 
low-energy states
(discussed in more detail in the next section). 
The critical behavior of 
the
DoS in the vicinity of zero energy 
is determined by
\begin{align}
	\nu(E)\propto |E|^{(2 - z)/z}.\label{dos}
\end{align}

In our numerical studies, the 
dynamic critical exponent is
extracted from the power-law behavior of the DoS in the chiral region
(as shown e.g.\ in
Fig.~\ref{DiracDoS}). 
Instead of calculating 
the
DoS directly, 
we first define\cite{Motrunich02} the 
quantity
	$N(E)=\sum_{j}\theta(E_j)\theta(E-E_j)$, 
where $j$ runs over the energy levels and $\theta(x)$ is the Heaviside step function. 
$N(E)$ is proportional to the DoS 
integrated
over $E$, which has a power law $E^{2/z}$ for 
$E \rightarrow 0$.

%%%%%%%%%%%%%%%%%%%%%%%%%%%%%%%%%%%%%%%%%%%%%%%%%%%%%%%%%%%%%%%%%%%%%%%%%%%%%%%%%%%%%%%%%
%%%%%%%%%%%%%%%%%%%%%%%%%%%%%%%%%%%%%%%%%%%%%%%%%%%%%%%%%%%%%%%%%%%%%%%%%%%%%%%%%%%%%%%%%
%%%%%%%%%%%%%%%%%%%%%%%%%%%%%%%%%%%%%%%%%%%%%%%%%%%%%%%%%%%%%%%%%%%%%%%%%%%%%%%%%%%%%%%%%
%%%%%%%%%%%%%%%%%%%%%%%%%%%%%%%%%%%%%%%%%%%%%%%%%%%%%%%%%%%%%%%%%%%%%%%%%%%%%%%%%%%%%%%%%
%%%%%%%%%%%%%%%%%%%%%%%%%%%%%%%%%%%%%%%%%%%%%%%%%%%%%%%%%%%%%%%%%%%%%%%%%%%%%%%%%%%%%%%%%
%%%%%%%%%%%%%%%%%%%%%%%%%%%%%%%%%%%%%%%%%%%%%%%%%%%%%%%%%%%%%%%%%%%%%%%%%%%%%%%%%%%%%%%%%
%%%%%%%%%%%%%%%%%%%%%%%%%%%%%%%%%%%%%%%%%%%%%%%%%%%%%%%%%%%%%%%%%%%%%%%%%%%%%%%%%%%%%%%%%
%%%%%%%%%%%%%%%%%%%%%%%%%%%%%%%%%%%%%%%%%%%%%%%%%%%%%%%%%%%%%%%%%%%%%%%%%%%%%%%%%%%%%%%%%

\subsection{DoS in MFD Approach \label{Sec: MFDDoS}}

\begin{figure}
\includegraphics[width=0.375\textwidth]{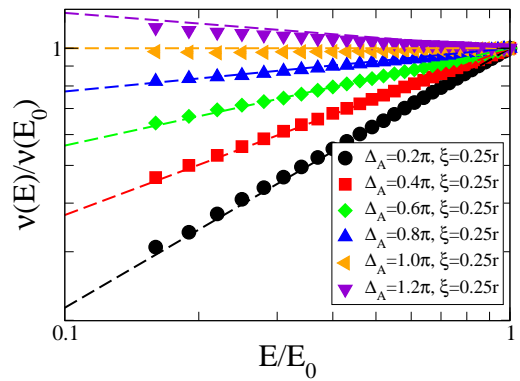}
\caption{The DoS in MFD near zero energy with $\mathcal{N}=40$, $\xi=0.25r$, 
and
$r=L/\mathcal{N}$. 
The results were obtained by averaging over 80 realizations of the disorder.
Dots are the numerical results from 200 energy levels, and the solid lines are the 
analytical formula.
From bottom to top: $\Delta_A=$ $0.2\pi$, $0.4\pi$, $0.6\pi$, $0.8\pi$, $\pi$, and $1.2\pi$. 
The data are rescaled so that the rightmost points are placed at the same position. 
As described in the text, for $\Delta_A \geq \pi$ we extract an effective disorder
strength from the data, which is later employed in the study of level statistics and 
Chalker scaling.
$\Delta_{A,\text{eff}}=0.96\pi$ for $\Delta_A=\pi$ and $\Delta_{A,\text{eff}}=1.125\pi$ for $\Delta_A=1.2\pi$.
\\[2mm]}
\label{DoSN40}
\end{figure}

In MFD approach, the white-noise correlation is replaced by a finite-ranged Gaussian distribution.
[See Eq.~(\ref{AParam}) and the discussion following.]
The DoS in the chiral region shows power-law behavior for a suitable choice of the 
Gaussian correlation length
$\xi$. In general, the DoS depends on $\Delta_A$, $\xi$, and 
the mode cutoff
$\mathcal{N}$. For a given $\Delta_A$ and $\mathcal{N}$, we choose a value of $\xi$ such 
that the power-law exponent 
reproduces the result in
Eq.~(\ref{dyn_exp}) for the single-valley model. In Fig.~\ref{DoSN40}, we present the 
power-law behavior of the DoS in this formalism. For $0<\Delta_A<\pi$ and $\mathcal{N}=$32, 40, 48, and 64, 
we are able to obtain the expected power law in Eq.~(\ref{dyn_exp}) with 
a fixed common value of the Gaussian correlation length
$\xi=0.25r$, where $r=L/\mathcal{N}$ is fixed.\cite{footnote--xival}
	
\begin{figure}
\includegraphics[width=0.375\textwidth]{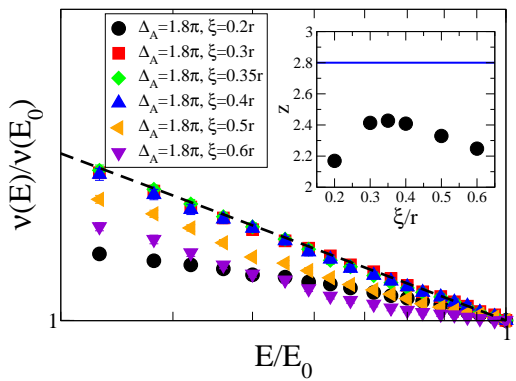}
\caption{The $\xi$-dependence of 
the
DoS in MFD. $\mathcal{N}=40$, $\Delta_A=1.8\pi$, $r=L/\mathcal{N}$, 
and 
we average over 20 disorder realizations.
Inset: The dynamical exponent $z$ 
computed
with different $\xi$
values.
The dots are extracted from the numerical DoS. 
The blue solid line is the analytical dynamical exponent with $\Delta_A=1.8\pi$. 
We
choose $\xi=0.35r$ as the proper parameter in MFD,
because this is where $z$ is least sensitive to the value of $\xi$.
We extract
$\Delta_{A,\text{eff}}\approx 1.42 \pi$.
}
\label{DoS_xi_MFD}
\end{figure}

For 
$\Delta_A \geq\pi$, the choice
$\xi=0.25r$ can no longer produce the expected power law. Instead of using the fixed value of 
$\xi$, we explore the $\xi$-dependence of the power law in the DoS. There is an intermediate 
region where the dependence 
of the DoS exponent
on $\xi$ is rather 
weak,
as exemplified in Fig.~\ref{DoS_xi_MFD}.
We extract the 
\emph{effective} dynamical exponent from this insensitive region,
and use Eq.~(\ref{dyn_exp}) to convert this to an 
\emph{effective disorder strength} 
$\Delta_{A,\text{eff}}$. In the sequel, we will use this effective disorder strength to
compare analytical and numerical results for level spacing statistics and Chalker scaling.
The dynamical exponents extracted for $\Delta_A>\pi$ are always smaller than the analytical 
predictions, 
so that $\Delta_{A,\text{eff}} < \Delta_A$.
We assume that the physics in the chiral region is governed by the effective 
disorder strength $\Delta_{A,\text{eff}}$
rather than the input value of $\Delta_A$.

%%%%%%%%%%%%%%%%%%%%%%%%%%%%%%%%%%%%%%%%%%%%%%%%%%%%%%%%%%%%%%%%%%%%%%%%%%%%%%%%%%%%%%%%%
%%%%%%%%%%%%%%%%%%%%%%%%%%%%%%%%%%%%%%%%%%%%%%%%%%%%%%%%%%%%%%%%%%%%%%%%%%%%%%%%%%%%%%%%%
%%%%%%%%%%%%%%%%%%%%%%%%%%%%%%%%%%%%%%%%%%%%%%%%%%%%%%%%%%%%%%%%%%%%%%%%%%%%%%%%%%%%%%%%%
%%%%%%%%%%%%%%%%%%%%%%%%%%%%%%%%%%%%%%%%%%%%%%%%%%%%%%%%%%%%%%%%%%%%%%%%%%%%%%%%%%%%%%%%%
%%%%%%%%%%%%%%%%%%%%%%%%%%%%%%%%%%%%%%%%%%%%%%%%%%%%%%%%%%%%%%%%%%%%%%%%%%%%%%%%%%%%%%%%%
%%%%%%%%%%%%%%%%%%%%%%%%%%%%%%%%%%%%%%%%%%%%%%%%%%%%%%%%%%%%%%%%%%%%%%%%%%%%%%%%%%%%%%%%%
%%%%%%%%%%%%%%%%%%%%%%%%%%%%%%%%%%%%%%%%%%%%%%%%%%%%%%%%%%%%%%%%%%%%%%%%%%%%%%%%%%%%%%%%%
%%%%%%%%%%%%%%%%%%%%%%%%%%%%%%%%%%%%%%%%%%%%%%%%%%%%%%%%%%%%%%%%%%%%%%%%%%%%%%%%%%%%%%%%%

\subsection{DoS in MDH Model}

\begin{figure}
 \centering
 \subfigure[$\pi$-flux lattice]{
\includegraphics[width=0.375\textwidth]{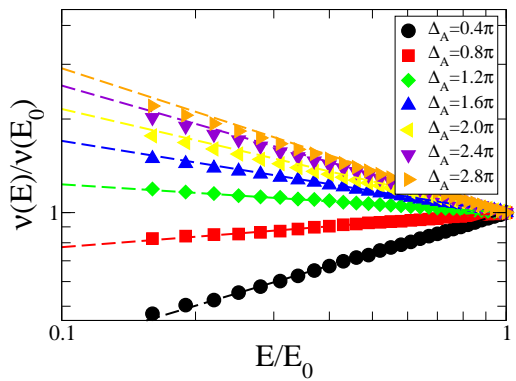}
   }
\\[8mm]
 \subfigure[Honeycomb lattice]{
\includegraphics[width=0.375\textwidth]{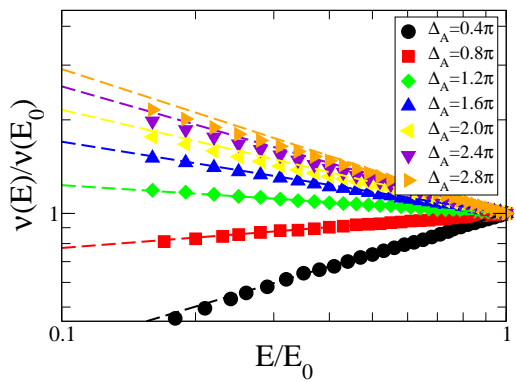}
   }
\caption{The DoS near zero energy for 
the
MDH model on 
(a) 
the
$\pi$-flux lattice with $L=256$ and 
(b) honeycomb lattice with the same size. 
Results are averaged over 40 disorder realizations.
Dots are the numerical results from 149 energy levels 
(excluding the first positive level), and the dashed lines are the 
analytical formula. The data are rescaled such that the rightmost 
points are placed at the same position.}
\label{DoSL256}
\end{figure}

The power-law behavior of the DoS in the MDH model has been reported previously.\cite{Motrunich02} 
We demonstrate the numerical results for $L=$256 in Fig. \ref{DoSL256} for 
the
$\pi$-flux 
square
and honeycomb lattices.

In the
weak disorder region $\Delta_A < 2\pi$, the dynamical exponents fit 
Eq.~(\ref{dyn_exp}). 
In the
strong disorder region $\Delta_A>2\pi$, the dynamical exponents start to show 
deviations from 
the
analytical formula. The deviations are due to finite size effects, 
for instance, finiteness of the mass 
terms.\cite{mass_term} The deviations in the power law are larger in the 
honeycomb lattice case. This is because the mass terms 
arise from
period-3 Kekul\'e patterns 
on the
honeycomb lattice, so the corresponding coarse 
graining cell needs to be at least a 6-site hexagon. For the $\pi$-flux lattice, 
the smallest coarse graining cell is a 2-by-2 block. 
For this reason, we expect that the
MDH model on 
the
honeycomb lattice will be more 
sensitive
finite size 
effects
than on the $\pi$-flux lattice.

%%%%%%%%%%%%%%%%%%%%%%%%%%%%%%%%%%%%%%%%%%%%%%%%%%%%%%%%%%%%%%%%%%%%%%%%%%%%%%%%%%%%%%%%%
%%%%%%%%%%%%%%%%%%%%%%%%%%%%%%%%%%%%%%%%%%%%%%%%%%%%%%%%%%%%%%%%%%%%%%%%%%%%%%%%%%%%%%%%%
%%%%%%%%%%%%%%%%%%%%%%%%%%%%%%%%%%%%%%%%%%%%%%%%%%%%%%%%%%%%%%%%%%%%%%%%%%%%%%%%%%%%%%%%%
%%%%%%%%%%%%%%%%%%%%%%%%%%%%%%%%%%%%%%%%%%%%%%%%%%%%%%%%%%%%%%%%%%%%%%%%%%%%%%%%%%%%%%%%%
%%%%%%%%%%%%%%%%%%%%%%%%%%%%%%%%%%%%%%%%%%%%%%%%%%%%%%%%%%%%%%%%%%%%%%%%%%%%%%%%%%%%%%%%%
%%%%%%%%%%%%%%%%%%%%%%%%%%%%%%%%%%%%%%%%%%%%%%%%%%%%%%%%%%%%%%%%%%%%%%%%%%%%%%%%%%%%%%%%%
%%%%%%%%%%%%%%%%%%%%%%%%%%%%%%%%%%%%%%%%%%%%%%%%%%%%%%%%%%%%%%%%%%%%%%%%%%%%%%%%%%%%%%%%%
%%%%%%%%%%%%%%%%%%%%%%%%%%%%%%%%%%%%%%%%%%%%%%%%%%%%%%%%%%%%%%%%%%%%%%%%%%%%%%%%%%%%%%%%%
%%%%%%%%%%%%%%%%%%%%%%%%%%%%%%%%%%%%%%%%%%%%%%%%%%%%%%%%%%%%%%%%%%%%%%%%%%%%%%%%%%%%%%%%%
%%%%%%%%%%%%%%%%%%%%%%%%%%%%%%%%%%%%%%%%%%%%%%%%%%%%%%%%%%%%%%%%%%%%%%%%%%%%%%%%%%%%%%%%%
%%%%%%%%%%%%%%%%%%%%%%%%%%%%%%%%%%%%%%%%%%%%%%%%%%%%%%%%%%%%%%%%%%%%%%%%%%%%%%%%%%%%%%%%%
%%%%%%%%%%%%%%%%%%%%%%%%%%%%%%%%%%%%%%%%%%%%%%%%%%%%%%%%%%%%%%%%%%%%%%%%%%%%%%%%%%%%%%%%%
%%%%%%%%%%%%%%%%%%%%%%%%%%%%%%%%%%%%%%%%%%%%%%%%%%%%%%%%%%%%%%%%%%%%%%%%%%%%%%%%%%%%%%%%%
%%%%%%%%%%%%%%%%%%%%%%%%%%%%%%%%%%%%%%%%%%%%%%%%%%%%%%%%%%%%%%%%%%%%%%%%%%%%%%%%%%%%%%%%%
%%%%%%%%%%%%%%%%%%%%%%%%%%%%%%%%%%%%%%%%%%%%%%%%%%%%%%%%%%%%%%%%%%%%%%%%%%%%%%%%%%%%%%%%%
%%%%%%%%%%%%%%%%%%%%%%%%%%%%%%%%%%%%%%%%%%%%%%%%%%%%%%%%%%%%%%%%%%%%%%%%%%%%%%%%%%%%%%%%%

\section{Multifractal Spectra \label{Sec: MFC}}

The zero-energy wavefunction of 
the
single-valley model in the continuum 
can be written down
explicitly\cite{Ludwig94} for a fixed disordered realization.
The exact multifractal spectrum\cite{Huckestein95,Evers08_RMP} 
is known for this state.\cite{Ludwig94,Castillo97,Carpentier01}
The multifractal spectrum 
measures
the statistics of 
the
local DoS, which can be measured experimentally
by scanning tunneling microscopy.\cite{Richardella10} 
It is also a useful tool to understand the characteristics of extended states in 
disordered environments. 
One defines
the inverse participation ratio (IPR), $P^{(q)}$
via
\begin{align}
	P^{(q)}(b,L)
	=
	\sum_{\vex{x}}|\psi_b(\vex{x})|^{2q}\propto \left(\frac{b}{L}\right)^{\tau(q)},
	\label{IPR}
\end{align}
where 
$|\psi_b(\bm{x})|^2$ corresponds to the probability of finding a particle in 
a box of size $b \ll L$
at position $\vex{x}$,
and $\tau(q)$ is the multifractal exponent associated with the $q$th IPR. 
$\tau(q)$ is a self-averaging quantity\cite{Chamon96} 
that satisfies the conditions
$\tau(1)=0$ 
due
to the normalization
and $\tau(0)=-d$.
The latter
reflects the 
dimension of the system, assuming a
system 
volume
$L^d$. 
The IPR satisfies the scaling form when $b$ is much larger than any microscopic scale 
and much smaller than the system size.\cite{Huckestein95,Evers08_RMP}

For a fixed system size $L$,
the multifractal exponents can be obtained by performing the numerical derivative 
of $\ln P^{(q)}(b,L)$ with 
respect to
different values of $b$. For example, the $\tau(q)$ for the plane wave is simply 
$d(q-1)$ because the probability of finding a particle is uniform. In the presence 
of 
disorder,
critically delocalized wavefunctions extend throughout the sample with an intricate 
inhomogeneous structure. 
For weak mulifractality and small $q$, the $\tau(q)$ can be approximated by
\begin{align}
	\tau(q) = d(q-1) - \theta q(q-1),
	\label{para}
\end{align}
where $\theta$ can be viewed as the degree of multifractality.

When $q$ exceeds a certain termination 
threshold\cite{Chamon96}
$q_{\text{c}}$, $\tau(q)$ becomes linearly proportional to $q$. 
$q_{\text{c}}$ specifies the region violating the parabolic approximation in 
Eq.~(\ref{para}). The multifractal spectrum for $q>q_c$ is governed by an 
extremum of the probability distribution,
and this is represented
by a single exponent rather than multiple fractal exponents.

The analytical $\tau(q)$ 
spectrum
for zero energy states shows non-analyticity at $\Delta_A=2\pi$. The 
$\tau(q)$
result\cite{Ludwig94,Chamon96,Castillo97,Carpentier01} 
for $\Delta_A\le 2\pi$ is
\begin{align}
	\tau(q)
	=
	\begin{cases}
		2\left(1-\frac{\Delta_A}{2\pi}q\right)(q-1), & 0\le q\le \sqrt{\frac{2\pi}{\Delta_A}},\\
		2\left(1-\sqrt{\frac{\Delta_A}{2\pi}}\right)^2q, & q>\sqrt{\frac{2\pi}{\Delta_A}}.
	\end{cases}
	\label{tau_w}
\end{align}
For $\Delta_A\ge 2\pi$,
\begin{align}
	\tau(q)
	=
	\begin{cases}
		-2\left(1-\sqrt{\frac{\Delta_A}{2\pi}}q\right)^2, & 0\le q\le \sqrt{\frac{2\pi}{\Delta_A}},\\
		0, & q>\sqrt{\frac{2\pi}{\Delta_A}}.
	\end{cases}\label{tau_s}
\end{align}
The 
termination threshold
$q_c=\sqrt{\frac{2\pi}{\Delta_A}}$ for both regions.

The zero-energy wavefunction shows 
``freezing'' behavior when $\Delta_A>2\pi$. 
The frozen wavefunction 
is
almost zero everywhere, except 
for
several 
well-localized
peaks with arbitrary separations.\cite{Chamon96,Castillo97,Carpentier01} 
It is qualitatively 
different from
the weakly multifractal extended states with $\Delta_A < 2 \pi$ that fill the sample
volume uniformly (but with an intricate structure of many peaks and 
valleys---see Fig.~\ref{Zero_Wavefcn}), and from
the usual localized state which is dominated by a single peak. 
The $\tau(q)$ for the frozen 
state 
is exactly zero
for $q > 1 \ge q_c$,
which is the same as a localized state.
The multifractal behavior can only be 
observed
for fractional 
values
of $q$.

A related quantity is the singularity spectrum\cite{Evers08_RMP} $f(\alpha)$, 
defined by the Legendre transformation of $\tau(q)$
\begin{align}
	f(\alpha)=\alpha q-
	\tau(q),
\end{align}
where $\alpha = d\tau/dq$.
The physical interpretation of $f(\alpha)$ is the following. 
Assume there is
a collection of points in position space 
where the
probability density $|\psi(\vex{x})|^2\propto L^{-\alpha}$. 
Then
the number of such points scales as $L^{f(\alpha)}$. For example, 
for a plane wave the $f(\alpha)$ spectrum is zero everywhere except $\alpha = d$. 
For a multifractal wavefunction, 
$f(\alpha)$ is a peaked function with non-zero width. 
The spectrum gets broader 
with increasing multifractality.

There are a handful of general properties regarding $f(\alpha)$. When 
$\alpha=\alpha_0 \equiv(d\tau/dq)|_{q=0}$, 
$f(\alpha_0)=d$ is maximized. 
When $\alpha = \alpha_1 \equiv(d\tau/dq)|_{q=1}$, $f(\alpha_1)=\alpha_1$ and $f'(\alpha_1)=1$.

The analytical $f(\alpha)$ for 
the
zero-energy wavefunction
is given by\cite{Chamon96,Castillo97}
\begin{align}
	f(\alpha)= 8 \frac{(d_+-\alpha)(\alpha-d_-)}{(d_+-d_-)^2}.
	\label{fa}
\end{align}
In the weak disorder regime $0\le\Delta_A<2\pi$,
\begin{align}
	d_{\pm}=2\left({\textstyle{1\pm\sqrt{\frac{\Delta_A}{2\pi}}}}\right)^2.
	\label{faw}
\end{align}
In the frozen phase $\Delta_A\ge 2\pi$,
\begin{align}
	d_-&=0,\,\,\,d_+=8\sqrt{\frac{\Delta_A}{2\pi}}.\label{fas}
\end{align}
$d_-=0$ indicates that $f(0)=0$. This is the signature of freezing in the $f(\alpha)$ spectrum.

In our simulations, we select the first positive energy state as representative. 
It is important to note that all the wavefunctions in the chiral region 
show similar multifractal characteristics reflecting the (effective) 
disorder strength dependence in the low-energy theory for both 
the
MFD and MDH models.

%%%%%%%%%%%%%%%%%%%%%%%%%%%%%%%%%%%%%%%%%%%%%%%%%%%%%%%%%%%%%%%%%%%%%%%%%%%%%%%%%%%%%%%%%
%%%%%%%%%%%%%%%%%%%%%%%%%%%%%%%%%%%%%%%%%%%%%%%%%%%%%%%%%%%%%%%%%%%%%%%%%%%%%%%%%%%%%%%%%
%%%%%%%%%%%%%%%%%%%%%%%%%%%%%%%%%%%%%%%%%%%%%%%%%%%%%%%%%%%%%%%%%%%%%%%%%%%%%%%%%%%%%%%%%
%%%%%%%%%%%%%%%%%%%%%%%%%%%%%%%%%%%%%%%%%%%%%%%%%%%%%%%%%%%%%%%%%%%%%%%%%%%%%%%%%%%%%%%%%
%%%%%%%%%%%%%%%%%%%%%%%%%%%%%%%%%%%%%%%%%%%%%%%%%%%%%%%%%%%%%%%%%%%%%%%%%%%%%%%%%%%%%%%%%
%%%%%%%%%%%%%%%%%%%%%%%%%%%%%%%%%%%%%%%%%%%%%%%%%%%%%%%%%%%%%%%%%%%%%%%%%%%%%%%%%%%%%%%%%
%%%%%%%%%%%%%%%%%%%%%%%%%%%%%%%%%%%%%%%%%%%%%%%%%%%%%%%%%%%%%%%%%%%%%%%%%%%%%%%%%%%%%%%%%
%%%%%%%%%%%%%%%%%%%%%%%%%%%%%%%%%%%%%%%%%%%%%%%%%%%%%%%%%%%%%%%%%%%%%%%%%%%%%%%%%%%%%%%%%

\subsection{Multifractal Spectra in MFD}

\begin{figure}
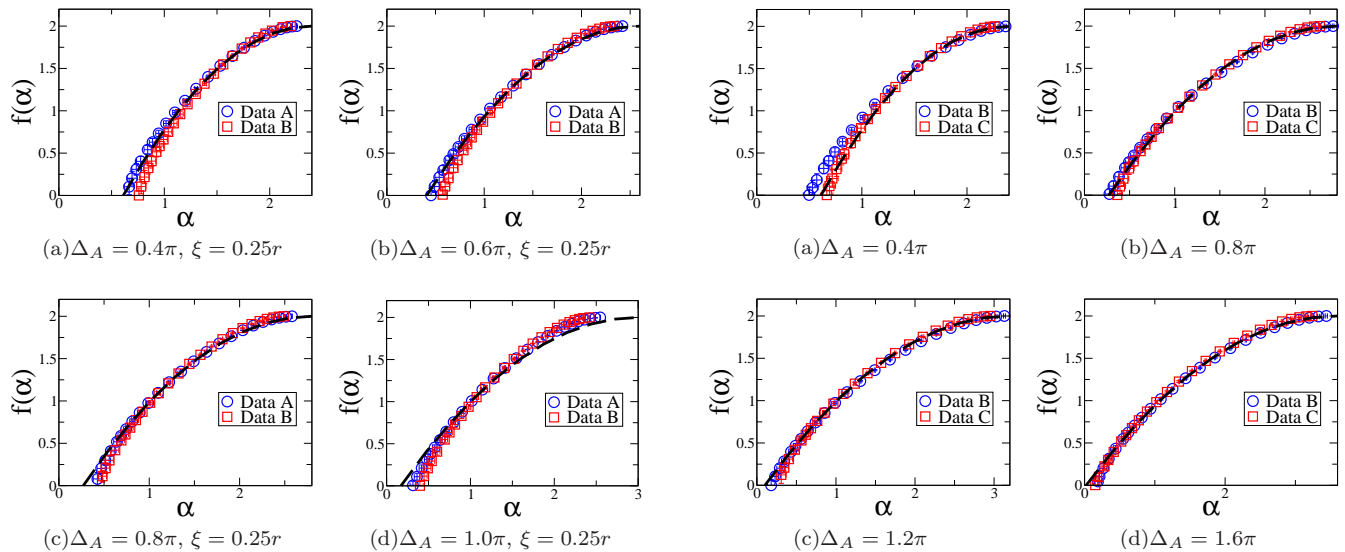

 \centering
 \subfigure[$\Delta_A=0.4\pi$, $\xi=0.25r$]{
\includegraphics[width=0.225\textwidth]{faN40_g04_xi025.eps}
   }
 \subfigure[$\Delta_A=0.6\pi$, $\xi=0.25r$]{
\includegraphics[width=0.225\textwidth]{faN40_g06_xi025.eps}
   }\\[3mm]
 \subfigure[$\Delta_A=0.8\pi$, $\xi=0.25r$]{
\includegraphics[width=0.225\textwidth]{faN40_g08_xi025.eps}
   }
 \subfigure[$\Delta_A=1.0\pi$, $\xi=0.25r$]{
\includegraphics[width=0.225\textwidth]{faN40_g1_xi025.eps}
   }
\caption{The $f(\alpha)$ spectra of low energy states in MFD with $\mathcal{N}=40$. 
For each 
value of $\Delta_A$, results are averaged over 40 disorder realizations.
Here $r=L/\mathcal{N}$. The data are extracted from the numerical derivative of 
the
IPR with respect to $b$, the size of the binning cell. Data A is extracted from $b=1$ and $b=2$. 
Data B is extracted from $b=2$ and $b=4$. The solid lines are the analytical prediction from 
Eqs.~(\ref{fa}) and (\ref{faw}).}
\label{faMFD}
\end{figure}

We first consider the results of our momentum space Dirac (MFD) calculations.
The multifractal spectra are consistent with 
Eqs.~(\ref{fa}) and (\ref{faw})
for $\Delta_A\le\pi$. These results are shown in Fig.~\ref{faMFD}.
For $\Delta_A=\pi$, the multifractal spectrum in MFD shows deviations from the analytical formulas. 
It is difficult to extract strong multifractal phenomena such as freezing using the MFD approach,
due to finite size limitations.
The finest spatial resolution in MFD is 
determined by the
$(2\mathcal{N}+1)$-by-$(2\mathcal{N}+1)$ grid. 
However, it contains some 
short-distance artifacts
due to the high momentum states ($|\vex{k}|>\Lambda$)
[see Fig.~\ref{DiracDoS}].
Instead, we convert our wavefunction in MFD to 
an $\mathcal{N}$-by-$\mathcal{N}$ grid.
Our grid sizes for MFD are 32-by-32, 40-by-40, 48-by-48, and 64-by-64. Calculating 
the
IPR in this formalism is restricted by 
$\mathcal{N}$
and the value of $\xi$. 
The wavefunctions with such small grid sizes can only represent weak multifractality. 

We find that states with
energy sufficiently away from the chiral region
show universal weak multifractality,
consistent with the
critical states of the integer quantum Hall plateau transition. 
We 
postpone
the discussion 
to Sec.~\ref{Sec: QHPM}.

%%%%%%%%%%%%%%%%%%%%%%%%%%%%%%%%%%%%%%%%%%%%%%%%%%%%%%%%%%%%%%%%%%%%%%%%%%%%%%%%%%%%%%%%%
%%%%%%%%%%%%%%%%%%%%%%%%%%%%%%%%%%%%%%%%%%%%%%%%%%%%%%%%%%%%%%%%%%%%%%%%%%%%%%%%%%%%%%%%%
%%%%%%%%%%%%%%%%%%%%%%%%%%%%%%%%%%%%%%%%%%%%%%%%%%%%%%%%%%%%%%%%%%%%%%%%%%%%%%%%%%%%%%%%%
%%%%%%%%%%%%%%%%%%%%%%%%%%%%%%%%%%%%%%%%%%%%%%%%%%%%%%%%%%%%%%%%%%%%%%%%%%%%%%%%%%%%%%%%%
%%%%%%%%%%%%%%%%%%%%%%%%%%%%%%%%%%%%%%%%%%%%%%%%%%%%%%%%%%%%%%%%%%%%%%%%%%%%%%%%%%%%%%%%%
%%%%%%%%%%%%%%%%%%%%%%%%%%%%%%%%%%%%%%%%%%%%%%%%%%%%%%%%%%%%%%%%%%%%%%%%%%%%%%%%%%%%%%%%%
%%%%%%%%%%%%%%%%%%%%%%%%%%%%%%%%%%%%%%%%%%%%%%%%%%%%%%%%%%%%%%%%%%%%%%%%%%%%%%%%%%%%%%%%%
%%%%%%%%%%%%%%%%%%%%%%%%%%%%%%%%%%%%%%%%%%%%%%%%%%%%%%%%%%%%%%%%%%%%%%%%%%%%%%%%%%%%%%%%%

\subsection{Multifractal Spectra in the MDH Model \label{Sec: MFC-MDH}}

\begin{figure}
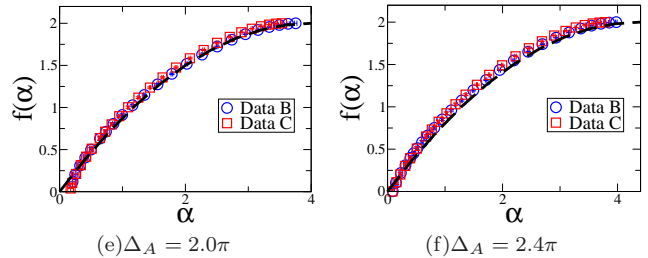

 \centering
 \subfigure[$\Delta_A=0.4\pi$]{
\includegraphics[width=0.225\textwidth]{faL256g04.eps}
   }
 \subfigure[$\Delta_A=0.8\pi$]{
\includegraphics[width=0.225\textwidth]{faL256g08.eps}
   }\\[3mm]
 \subfigure[$\Delta_A=1.2\pi$]{
\includegraphics[width=0.225\textwidth]{faL256g1p2.eps}
   }
 \subfigure[$\Delta_A=1.6\pi$]{
\includegraphics[width=0.225\textwidth]{faL256g1p6.eps}
   }\\[3mm]
 \subfigure[$\Delta_A=2.0\pi$]{
\includegraphics[width=0.225\textwidth]{faL256g2.eps}
   }
 \subfigure[$\Delta_A=2.4\pi$]{
\includegraphics[width=0.225\textwidth]{faL256g2p4.eps}
   }
 \caption{The $f(\alpha)$ spectra of 
low
energy states of the MDH model with $L=256$. 
For each 
value of $\Delta_A$, results are averaged over 40 disorder realizations.
The data are extracted from the numerical derivative of 
the
IPR with respect to $b$, the size of the binning cell. Data B is extracted from 
$b=2$ and $b=4$. Data C is extracted from $b=4$ and $b=8$. 
The solid lines are the analytical prediction from Eqs.~(\ref{fa}), (\ref{faw}), and (\ref{fas}).}
\label{MFSL256}
\end{figure}

In order to simulate 
Dirac 
fermions coupled only to vector potential disorder
with the MDH model, one has to perform 
a
coarse graining (binning) procedure. 
For the square lattice with $\pi$-flux,
the binning size $b$ needs to be at least twice larger than the 
lattice constant (corresponding to a 2-by-2 coarse graining cell). 
We calculate the multifractal spectrum for $L=$128, 256, and 512. The finite size 
scaling of the single-valley model contains $1/\ln L$ and $\ln\ln L/\ln L$ terms.\cite{Carpentier01}
In this aspect, it is difficult to obtain reliable finite size scaling. 
The $f(\alpha)$ spectra for different sizes are almost identical in our simulations. 
We only present multifractal spectra with $L=256$ in the Fig.~\ref{MFSL256}. 
The results fit the analytical 
formula for
$f(\alpha)$ 
in Eq.~(\ref{fa}), and in particular
reveal 
the signature $f(0) = 0$ for the frozen regime with
$\Delta_A=2\pi$ and $2.4\pi$. 

The finite energy states of 
the
MDH model are expected to be localized in the thermodynamic limit.\cite{Motrunich02} 
Only the states in the chiral region reflect the physics of the Dirac fermion with 
the
random vector potential.

%%%%%%%%%%%%%%%%%%%%%%%%%%%%%%%%%%%%%%%%%%%%%%%%%%%%%%%%%%%%%%%%%%%%%%%%%%%%%%%%%%%%%%%%%
%%%%%%%%%%%%%%%%%%%%%%%%%%%%%%%%%%%%%%%%%%%%%%%%%%%%%%%%%%%%%%%%%%%%%%%%%%%%%%%%%%%%%%%%%
%%%%%%%%%%%%%%%%%%%%%%%%%%%%%%%%%%%%%%%%%%%%%%%%%%%%%%%%%%%%%%%%%%%%%%%%%%%%%%%%%%%%%%%%%
%%%%%%%%%%%%%%%%%%%%%%%%%%%%%%%%%%%%%%%%%%%%%%%%%%%%%%%%%%%%%%%%%%%%%%%%%%%%%%%%%%%%%%%%%
%%%%%%%%%%%%%%%%%%%%%%%%%%%%%%%%%%%%%%%%%%%%%%%%%%%%%%%%%%%%%%%%%%%%%%%%%%%%%%%%%%%%%%%%%
%%%%%%%%%%%%%%%%%%%%%%%%%%%%%%%%%%%%%%%%%%%%%%%%%%%%%%%%%%%%%%%%%%%%%%%%%%%%%%%%%%%%%%%%%
%%%%%%%%%%%%%%%%%%%%%%%%%%%%%%%%%%%%%%%%%%%%%%%%%%%%%%%%%%%%%%%%%%%%%%%%%%%%%%%%%%%%%%%%%
%%%%%%%%%%%%%%%%%%%%%%%%%%%%%%%%%%%%%%%%%%%%%%%%%%%%%%%%%%%%%%%%%%%%%%%%%%%%%%%%%%%%%%%%%
%%%%%%%%%%%%%%%%%%%%%%%%%%%%%%%%%%%%%%%%%%%%%%%%%%%%%%%%%%%%%%%%%%%%%%%%%%%%%%%%%%%%%%%%%
%%%%%%%%%%%%%%%%%%%%%%%%%%%%%%%%%%%%%%%%%%%%%%%%%%%%%%%%%%%%%%%%%%%%%%%%%%%%%%%%%%%%%%%%%
%%%%%%%%%%%%%%%%%%%%%%%%%%%%%%%%%%%%%%%%%%%%%%%%%%%%%%%%%%%%%%%%%%%%%%%%%%%%%%%%%%%%%%%%%
%%%%%%%%%%%%%%%%%%%%%%%%%%%%%%%%%%%%%%%%%%%%%%%%%%%%%%%%%%%%%%%%%%%%%%%%%%%%%%%%%%%%%%%%%
%%%%%%%%%%%%%%%%%%%%%%%%%%%%%%%%%%%%%%%%%%%%%%%%%%%%%%%%%%%%%%%%%%%%%%%%%%%%%%%%%%%%%%%%%
%%%%%%%%%%%%%%%%%%%%%%%%%%%%%%%%%%%%%%%%%%%%%%%%%%%%%%%%%%%%%%%%%%%%%%%%%%%%%%%%%%%%%%%%%
%%%%%%%%%%%%%%%%%%%%%%%%%%%%%%%%%%%%%%%%%%%%%%%%%%%%%%%%%%%%%%%%%%%%%%%%%%%%%%%%%%%%%%%%%
%%%%%%%%%%%%%%%%%%%%%%%%%%%%%%%%%%%%%%%%%%%%%%%%%%%%%%%%%%%%%%%%%%%%%%%%%%%%%%%%%%%%%%%%%

\section{Level statistics and Chalker scaling in the single valley model \label{Sec: DynProp}}

The exact zero energy wavefunction in the single-valley model
has been extensively studied.\cite{Ludwig94,Chamon96,Castillo97,Hatsugai97,Carpentier01,Chen12} 
Besides the global density of states,\cite{Morita97,Motrunich02,Horovitz02,Mudry03}
the properties of low-energy states have received less attention.
We focus on two quantities related to correlations between states at different energies: the level 
spacing distribution and the two-wavefunction correlation. The former 
measures the distribution of gaps between nearby levels. It is also 
a
useful probe for Anderson localization. 
On the other hand, 
the
two-wavefunction correlation function characterizes the overlap 
of the probability distributions for
two wavefunctions at different energies. 
In this section, we numerically study the level spacing distribution and the two-wavefunction correlation 
in
the
MFD and MDH models.
We show that states at different energies are strongly correlated in the chiral region,
for both weak and strong disorder. Our main conclusion is that the spectral characteristics
discussed here do not exhibit clear signatures of the freezing transition observed in 
multifractal spectra and in the density of states.

%%%%%%%%%%%%%%%%%%%%%%%%%%%%%%%%%%%%%%%%%%%%%%%%%%%%%%%%%%%%%%%%%%%%%%%%%%%%%%%%%%%%%%%%%
%%%%%%%%%%%%%%%%%%%%%%%%%%%%%%%%%%%%%%%%%%%%%%%%%%%%%%%%%%%%%%%%%%%%%%%%%%%%%%%%%%%%%%%%%
%%%%%%%%%%%%%%%%%%%%%%%%%%%%%%%%%%%%%%%%%%%%%%%%%%%%%%%%%%%%%%%%%%%%%%%%%%%%%%%%%%%%%%%%%
%%%%%%%%%%%%%%%%%%%%%%%%%%%%%%%%%%%%%%%%%%%%%%%%%%%%%%%%%%%%%%%%%%%%%%%%%%%%%%%%%%%%%%%%%
%%%%%%%%%%%%%%%%%%%%%%%%%%%%%%%%%%%%%%%%%%%%%%%%%%%%%%%%%%%%%%%%%%%%%%%%%%%%%%%%%%%%%%%%%
%%%%%%%%%%%%%%%%%%%%%%%%%%%%%%%%%%%%%%%%%%%%%%%%%%%%%%%%%%%%%%%%%%%%%%%%%%%%%%%%%%%%%%%%%
%%%%%%%%%%%%%%%%%%%%%%%%%%%%%%%%%%%%%%%%%%%%%%%%%%%%%%%%%%%%%%%%%%%%%%%%%%%%%%%%%%%%%%%%%
%%%%%%%%%%%%%%%%%%%%%%%%%%%%%%%%%%%%%%%%%%%%%%%%%%%%%%%%%%%%%%%%%%%%%%%%%%%%%%%%%%%%%%%%%

\subsection{Level Spacing Distribution}

In a random quantum system, one can view the exact level spectrum in a fixed 
realization of the disorder as arising through the perturbative sewing together 
of spatially segregated subsystems. 
In a metallic phase, nearby energy levels repel each other.\cite{Sivan87}
States avoid level crossings due to the finite overlap of 
their spatial distributions.
By contrast, in an Anderson insulator,
different
states can be arbitrarily close in energy because the
spatial overlap of their probability densities 
is essentially zero.
The distribution of energy levels 
therefore
reflects
the localization properties of the 
phase.\cite{Mirlin00}

In the single-valley Dirac model, a representative wavefunction in the frozen regime\cite{Chamon96} 
that occurs for strong disorder $(\Delta_A>2\pi)$ typically possesses 
rare peaks with arbitrarily large separation between them.\cite{Ryu01A,Carpentier01} 
These states appear ``quasi-localized,'' as indicated by the vanishing multifractal spectrum 
$\tau(q)$ for $q > 1$ [Eq.~(\ref{tau_s})]; 
see also Fig.~\ref{Zero_Wavefcn}.
We might expect that the level spacing
distribution will reflect this, i.e.\ show Poissonian, rather than Wigner-Dyson statistics.
On the other hand, states at weak disorder are weakly multifractal and extended.
In fact,
our results 
show 
no signature 
of the freezing transition 
in 
the 
level spacing distribution.
In both 
 the 
MFD and MDH 
 models, 
the 
distributions are 
essentially 
independent of the disorder strength, 
 and are well-approximated by the Wigner surmise in the \emph{host} model
at non-zero energies. 

We first define the level spacing distribution function 
$P(s)$, which
satisfies 
\begin{align}
	\int_0^{\infty}P(s) ds=1, \;\;
	\int_0^{\infty}s P(s) ds=1\label{LS},
\end{align}
where 
$s=|E_n-E_{n+1}|/\delta(E_n)$
is the normalized level spacing. 
Here 
$\delta (E_n)$ is the mean level spacing 
near energy $E_n$. Diffusive metals in the 
Wigner-Dyson 
symmetry classes\cite{Mirlin00}  
can be described by the Wigner surmise $P(s)=As^{\beta}\exp(-Bs^2)$, where $A$ and $B$ 
are 
determined by 
Eq.~(\ref{LS}).
The parameter $\beta = \{1,2,4\}$ in the orthogonal, unitary, and symplectic classes,
respectively.
For localized states, the distribution is 
Poissonian
$P(s)=e^{-s}$.

In the single-valley Dirac problem, the DoS $\nu(E)$ changes rapidly in the low-energy 
chiral region; see Eqs.~(\ref{dyn_exp}) and (\ref{dos}), and Figs.~\ref{DiracDoS}, \ref{DoSN40}, and \ref{DoSL256}.
For both of the numerical 
 MFD and MDH 
approaches, 
we define the	
level distribution function 
by 
rescaling energy level intervals
relative to the local mean spacing $\delta(E) \propto 1/\nu(E)$.\cite{Evangelou03}

\begin{figure}
\includegraphics[width=0.4\textwidth]{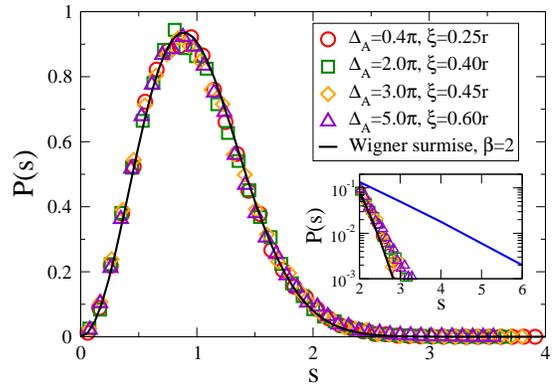}
\caption{The level spacing distribution in the chiral region of the Dirac fermion in MFD. 
Here the system size is 
$\mathcal{N}=32$, 
 we keep 
100 energy levels per realization,
 and we have averaged over
400 disorder realizations;
$r=L/\mathcal{N}$. The effective disorder strength 
$\Delta_{A,\text{eff}}$ for the 
 presented data is 
$0.4\pi$, $1.55\pi$, $1.9\pi$, and $2.55\pi$. Inset: The tail distribution. 
The	
black solid line is the Wigner surmise with $\beta=2$; 
the
blue solid line 
 is the Poisson	
distribution.\\[2mm]}
\label{DiracLS}
\end{figure}

In 
the	
MFD approach, 
we find that $P(s)$ 
in the chiral region 
fits the Wigner surmise with $\beta=2$ (unitary metal) 
for all the 
disorder strengths we explored, 
$0.4\pi\le\Delta_{A,\text{eff}}\le 2.55\pi$ (see Fig. \ref{DiracLS}). 
The distributions are independent of $\Delta_{A,\text{eff}}$.
(The procedure used to define the effective disorder strength 
$\Delta_{A,\text{eff}}$ was explained in Sec.~\ref{Sec: MFDDoS}.)

\begin{figure}
\includegraphics[width=0.4\textwidth]{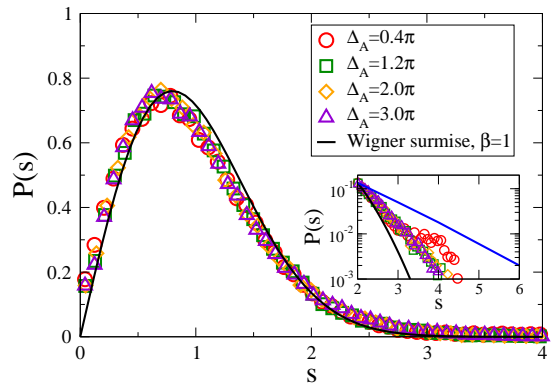}
\caption{The level spacing distribution in the chiral region of 
the	MDH model. 
The system size is $L=256$, 
we keep 149 energy levels per realization,
and we have averaged over 400 disorder realizations.	
Inset: The tail distribution. 
The 
black solid line is the Wigner surmise with $\beta=1$; 
the	
blue solid line 
 is the Poisson	
distribution.\\[2mm]}
\label{MDHLS}
\end{figure}

In 
the
MDH model, 
$P(s)$ also exhibits
level repulsion,
as shown in 
Fig. \ref{MDHLS}.
We exclude the first energy
interval	
because the first two 
levels are degenerate when $L\rightarrow\infty$.
The 
results	
are close to the Wigner surmise with $\beta=1$ (orthogonal metal) rather than $\beta=2$. 
There are 
deviations from 
the Wigner surmise 
(particularly in the tail),
consistent with 
a 
previous report.\cite{Morita97}
In the limit $L\rightarrow\infty$, the finite-energy states in the MDH model are always localized. 
The levels we sampled are in the chiral 
low-energy region, and reveal the same critical	
properties 
(dynamic critical exponent, multifractal spectra) as
the single-valley 
Dirac
model. 
For states in the MDH model sufficiently away from the chiral region, the 
level spacing distribution is Poissonian, which indicates localization. 

The 
results for the MFD and MDH models suggest that the level statistics in the chiral region are 
independent of the disorder strength.
At finite energy, the character of the (de)localization problem in these models is the same as 
that obtained by adding a non-zero chemical potential to the single particle Hamiltonian. 
This breaks the special chiral symmetry [Eq.~(\ref{chC})], reducing the system to one 
of the standard Wigner-Dyson classes. The single Dirac fermion model crosses over to the unitary 
class at finite energy (MFD approach), while the MDH lattice model crosses over to the orthogonal class.
Evidently the level statistics for these states reflect only the symmetry class of the ``host'' 
model at finite energy. In particular, $P(s)$ shows no signs of the freezing transition in the MDH model,
despite the fact that we do observe signatures in the DoS and multifractal spectrum (Figs.~\ref{DoSL256} and \ref{MFSL256}).
The results imply that the overlap of probabiltiy densities associated with different wavefunctions
is non-negligible, even for states in the frozen regime.

%%%%%%%%%%%%%%%%%%%%%%%%%%%%%%%%%%%%%%%%%%%%%%%%%%%%%%%%%%%%%%%%%%%%%%%%%%%%%%%%%%%%%%%%%
%%%%%%%%%%%%%%%%%%%%%%%%%%%%%%%%%%%%%%%%%%%%%%%%%%%%%%%%%%%%%%%%%%%%%%%%%%%%%%%%%%%%%%%%%
%%%%%%%%%%%%%%%%%%%%%%%%%%%%%%%%%%%%%%%%%%%%%%%%%%%%%%%%%%%%%%%%%%%%%%%%%%%%%%%%%%%%%%%%%
%%%%%%%%%%%%%%%%%%%%%%%%%%%%%%%%%%%%%%%%%%%%%%%%%%%%%%%%%%%%%%%%%%%%%%%%%%%%%%%%%%%%%%%%%
%%%%%%%%%%%%%%%%%%%%%%%%%%%%%%%%%%%%%%%%%%%%%%%%%%%%%%%%%%%%%%%%%%%%%%%%%%%%%%%%%%%%%%%%%
%%%%%%%%%%%%%%%%%%%%%%%%%%%%%%%%%%%%%%%%%%%%%%%%%%%%%%%%%%%%%%%%%%%%%%%%%%%%%%%%%%%%%%%%%
%%%%%%%%%%%%%%%%%%%%%%%%%%%%%%%%%%%%%%%%%%%%%%%%%%%%%%%%%%%%%%%%%%%%%%%%%%%%%%%%%%%%%%%%%
%%%%%%%%%%%%%%%%%%%%%%%%%%%%%%%%%%%%%%%%%%%%%%%%%%%%%%%%%%%%%%%%%%%%%%%%%%%%%%%%%%%%%%%%%

\subsection{Two-Eigenfunction Correlation}

To further characterize effects of weak and strong disorder, 
we 
compute the correlations between two wavefunctions at different energies 
in the same disorder 
realization.\cite{Chalker88,Chalker1990253,Huckestein94,Pracz96,Fyodorov97,Cuevas07,Kravtsov10}	
The correlation function in $d$
spatial dimensions
is defined by
\begin{align}\label{CEDef}
	C(E_0,E_0 + \epsilon; L)
	=
	\int d^d{\vex{x}}\,
	|\psi_{0}(\vex{x})|^2
	|\psi_{\epsilon}(\vex{x})|^2,
\end{align}
where $E_0$, $E_0+\epsilon$ are 
eigenenergies	
of the system, and 
$\psi_{0}$, $\psi_{\epsilon}$ are the corresponding wavefunctions. 
$C(E_0,E_0 + \epsilon;L)$ reduces to the inverse participation ratio (IPR) 
[Eq. (\ref{IPR})] with $q=2$ when $\epsilon=0$. 

This	
correlation function shows different behavior 
when evaluated in a region of extended or localized states.	
In particular, $C(E_0,E_0+\epsilon;L) \sim 0$ for localized states with 
$0 < \epsilon \ll \delta_l$, where $\delta_l$ is the level spacing in a
characteristic localization volume. This result obtains because states with 
nearby energies are typically separated in real space, so that the 
probability densities of the two wavefunctions have negligble overlap
for all $\vex{x}$. On the other hand, for states near a mobility edge,
$C(E_0,E_0+\epsilon;L)$ shows non-trivial scaling\cite{Chalker88,Chalker1990253,Fyodorov97,Cuevas07} 
in $\epsilon$.
To simplify notation,
we suppress the argument $E_0$ in the later discussion,
$C(E_0,E_0+\epsilon;L)\equiv C(\epsilon;L)$.

In order to understand the scaling behavior of $C$, we define 
\begin{align}
	F(\epsilon;L)
	\equiv
	\frac{\int d^d{\vex{x}}\,
	|\psi_{0}(\vex{x})|^2
	|\psi_{\epsilon}(\vex{x})|^2
	}{
	\int d^d{\vex{x}}\,|\psi_{0}(\vex{x})|^4}.
\end{align}
The general scaling form is
\begin{align}
	F(\epsilon;L)
	=
	\left(\frac{a}{L}\right)^{\delta}f(\epsilon L^z),
\end{align}
where 
$\delta$ is some scaling exponent
and $a$ represents a short 
distance 
scale. 
The exponent	 
$\delta$ must be zero because $F(0;L)$ is normalized 
to unity. We assume that $f(x)\sim x^{-\mu}$ for large $x$, 
which implies that
\begin{align}
	\lim_{L \rightarrow\infty}	
	C(\epsilon;L)
	\sim
	\frac{\epsilon^{-\mu}}{L^{d_2+\mu z}},\label{CS_C}
\end{align}
where 
$d_2=\tau(2)$	
is the 
correlation 
dimension.
On the other hand, the scaling behavior for large $\epsilon$ should be 
determined by integration over 
the product of 
the two eigenstate	
probability densities, 
instead of the second IPR. This implies that	
\begin{align}
	\mu
	=
	\frac{d - d_2}{z},
	\label{CS_mu}
\end{align}
where $d$ is the spatial dimension.
The 
result	
in Eq.~(\ref{CS_mu}) generalizes the well-known Chalker scaling exponent\cite{Chalker88,Chalker1990253} 
to	
a system with a critical low-energy DoS ($z\neq d$). 

When $E_0=0$
in Eq.~(\ref{CEDef}),
the disorder dependent formula for $\mu$ is
\begin{align}
	\mu=\begin{cases}
	\frac{2\Delta_A/\pi}{1+\Delta_A/\pi}, &0\le\Delta_A\le\frac{\pi}{2},\\[2mm]
	\frac{2-4\left(1-\sqrt{\Delta_A/(2\pi)}\right)^2}{1+\Delta_A/\pi}, & \frac{\pi}{2}<\Delta_A\le 2\pi,\\[2mm]
	\frac{2}{4\sqrt{\Delta_A/(2\pi)}-1}, & 2\pi<\Delta_A.
	\end{cases}\label{CSansatz}
\end{align}
There are three regimes of the exponent $\mu$. 
The multifractal dimension
$d_2=\tau(2)$ has two non-analyticities at $\Delta_A=\pi/2$ and $\Delta_A=2\pi$; 
the dynamical exponent $z$ has a transition at $\Delta_A=2\pi$. For $\Delta_A<\pi/2$, $\mu$ is monotonically increasing 
and can be determined by 
the first expression in each of 
Eqs.~(\ref{dyn_exp}) and (\ref{tau_w}). When $\Delta_A$ is larger than $\pi/2$ 
($q_c = \sqrt{2 \pi / \Delta_a} < 2 $),	
one needs to apply the formula for termination in Eq.~(\ref{tau_w}). 
In the	
frozen regime
$\Delta_A > 2 \pi$,	
$d_2=0$ and the dynamical exponent is 
given by the second result in 
Eq.~(\ref{dyn_exp}).

\begin{figure}
\includegraphics[width=0.4\textwidth]{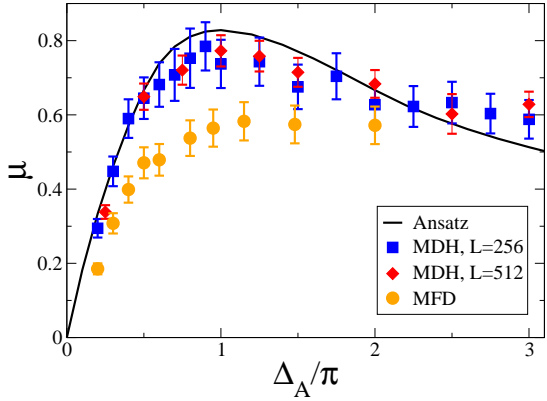}
\caption{The Chalker scaling exponent
$\mu$ [defined via Eq.~(\ref{CS_C})]
as a function of 
the 
disorder strength 
 in the 
MFD and MDH approaches. 
$\mathcal{N}=40$ and 
we average over 80 realizations of the disorder 
for MFD. 
We show data for two system sizes of the
MDH model. 
For 
$L=256$, 
we average over 200 realizations of the disorder.
For 
$L=512$, we average over 40 disorder realizations.
For MFD, the effective disorder strengths 
are presented when $\Delta_A>\pi$. 
For 
the 
MDH model, 
the wavefunctions are 
coarse-grained with binning size $b=2$.
The solid curve is the analytical prediction 
that includes termination and freezing effects,
Eq.~(\ref{CSansatz}).
}
\label{CS_exp}
\end{figure}

We calculate the 
disorder-averaged
$C(\epsilon;L)$ for $E_0=0$ in the chiral region for 
both
the
 MFD and MDH 
models.
The numerical exponent
shown in Fig.~\ref{CS_exp} is 
qualitatively consistent with 
generalized Chalker scaling [Eq.~(\ref{CSansatz})]
for 
weak disorder in MFD  
and 
for disorder strengths up to and beyond the freezing transition 
($\Delta_A\le 3\pi$) in 
the
MDH model. 
For the MFD calculations, we plot $\mu$ versus the 
effective disorder strength
$\Delta_{A,\text{eff}}$ for $\Delta_A>\pi$, as 
defined in Sec.~\ref{Sec: MFDDoS}.
The good agreement of the MDH model numerics with the analytical
prediction indicates the presence of strong correlations between the
probability density profiles (peaks and valleys) of different eigenstates,
for both weak and strong disorder. We conclude that while individual 
wavefunctions are strongly inhomogeneous in space in the frozen regime, 
quantum critical scaling survives---the spectral characteristics remain 
``ergodic.''

The 
discrepancy in 
the
MFD 
result for the generalized Chalker scaling exponent $\mu$
might come from 
finite system size limitations to
this approach. Similar to the situation for multifractal spectra, 
a high resolution 
is essential to extract the correct correlations
from the critical wavefunctions.
For 
the
MDH model, we perform the coarse graining procedure 
described in Sec.~\ref{Sec: MFC-MDH}
to the wavefunctions with binning size $b=2$.

%%%%%%%%%%%%%%%%%%%%%%%%%%%%%%%%%%%%%%%%%%%%%%%%%%%%%%%%%%%%%%%%%%%%%%%%%%%%%%%%%%%%%%%%%
%%%%%%%%%%%%%%%%%%%%%%%%%%%%%%%%%%%%%%%%%%%%%%%%%%%%%%%%%%%%%%%%%%%%%%%%%%%%%%%%%%%%%%%%%
%%%%%%%%%%%%%%%%%%%%%%%%%%%%%%%%%%%%%%%%%%%%%%%%%%%%%%%%%%%%%%%%%%%%%%%%%%%%%%%%%%%%%%%%%
%%%%%%%%%%%%%%%%%%%%%%%%%%%%%%%%%%%%%%%%%%%%%%%%%%%%%%%%%%%%%%%%%%%%%%%%%%%%%%%%%%%%%%%%%
%%%%%%%%%%%%%%%%%%%%%%%%%%%%%%%%%%%%%%%%%%%%%%%%%%%%%%%%%%%%%%%%%%%%%%%%%%%%%%%%%%%%%%%%%
%%%%%%%%%%%%%%%%%%%%%%%%%%%%%%%%%%%%%%%%%%%%%%%%%%%%%%%%%%%%%%%%%%%%%%%%%%%%%%%%%%%%%%%%%
%%%%%%%%%%%%%%%%%%%%%%%%%%%%%%%%%%%%%%%%%%%%%%%%%%%%%%%%%%%%%%%%%%%%%%%%%%%%%%%%%%%%%%%%%
%%%%%%%%%%%%%%%%%%%%%%%%%%%%%%%%%%%%%%%%%%%%%%%%%%%%%%%%%%%%%%%%%%%%%%%%%%%%%%%%%%%%%%%%%
%%%%%%%%%%%%%%%%%%%%%%%%%%%%%%%%%%%%%%%%%%%%%%%%%%%%%%%%%%%%%%%%%%%%%%%%%%%%%%%%%%%%%%%%%
%%%%%%%%%%%%%%%%%%%%%%%%%%%%%%%%%%%%%%%%%%%%%%%%%%%%%%%%%%%%%%%%%%%%%%%%%%%%%%%%%%%%%%%%%
%%%%%%%%%%%%%%%%%%%%%%%%%%%%%%%%%%%%%%%%%%%%%%%%%%%%%%%%%%%%%%%%%%%%%%%%%%%%%%%%%%%%%%%%%
%%%%%%%%%%%%%%%%%%%%%%%%%%%%%%%%%%%%%%%%%%%%%%%%%%%%%%%%%%%%%%%%%%%%%%%%%%%%%%%%%%%%%%%%%
%%%%%%%%%%%%%%%%%%%%%%%%%%%%%%%%%%%%%%%%%%%%%%%%%%%%%%%%%%%%%%%%%%%%%%%%%%%%%%%%%%%%%%%%%
%%%%%%%%%%%%%%%%%%%%%%%%%%%%%%%%%%%%%%%%%%%%%%%%%%%%%%%%%%%%%%%%%%%%%%%%%%%%%%%%%%%%%%%%%
%%%%%%%%%%%%%%%%%%%%%%%%%%%%%%%%%%%%%%%%%%%%%%%%%%%%%%%%%%%%%%%%%%%%%%%%%%%%%%%%%%%%%%%%%
%%%%%%%%%%%%%%%%%%%%%%%%%%%%%%%%%%%%%%%%%%%%%%%%%%%%%%%%%%%%%%%%%%%%%%%%%%%%%%%%%%%%%%%%%

\section{Quantum Hall Critical Metal: Finite Energy States \label{Sec: QHPM}}

In this section, we discuss the finite energy 
states
of the single-valley 
Dirac 
model. 
These 
belong to the unitary class (class A).\cite{Ludwig94} In two 
dimensions, 
this class is always localized except 
in 
the presence of 
topological protection. 
The finite energy physics of the single-valley 
model with vector potential disorder is expected
to be	
the same as 
that of the low-energy states for a single 
Dirac fermion subject to any combination of 
zero-mean mass, scalar, or vector disorder potentials\cite{Ludwig94} 
(i.e., at least two types with non-zero variance).
The states are expected to be critically delocalized at 
all energies, with critical properties governed by the 
plateau transition of the integer quantum Hall 
effect.\cite{Ludwig94,Nomura07_QHE,Ostrovsky07}

\begin{figure}[ht]
 \centering
 \subfigure[Weak disorder]{
\includegraphics[width=0.325\textwidth]{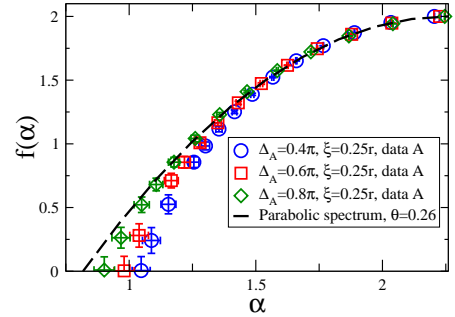}
   }\\[5mm]
 \subfigure[Intermediate disorder]{
\includegraphics[width=0.325\textwidth]{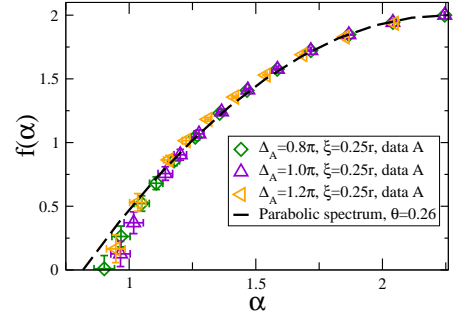}
   }
\caption{The $f(\alpha)$ spectra of finite energy states 
in the single-valley Dirac model using the 
momentum space formalism with $\mathcal{N}=40$. We perform 
an average over 80 realizations of the disorder;
$r=L/\mathcal{N}$. Finite energy states with $\Delta_A=0.4\pi$ 
and $\Delta_A=0.6\pi$ show deviations from the parabolic spectrum 
in Eq.~(\ref{fpara})
with 
$\theta=0.26$. 
The latter is a good approximation to the integer quantum Hall plateau transition spectrum.\cite{Evers01,Evers08,Obuse08}
For $\Delta_A=0.8\pi$, $\pi$, and $1.2\pi$, the $f(\alpha)$ spectra are 
consistent with the $\theta=0.26$ curve.
$\Delta_{A,\text{eff}}=0.96\pi$ for $\Delta_A=\pi$, and $\Delta_{A,\text{eff}}=1.125\pi$ for $\Delta_A=1.2\pi$. 
Data A is extracted from $b=1$ and $b=2$.\\[2mm]}
\label{FEMFS}
\end{figure}

We sample states around energy $\sim 0.6\Lambda$ with $\mathcal{N}=$32, 40, 48, and 64 in MFD, 
where $\Lambda = \mathcal{N}(2\pi/L)$ is the energy cutoff. 
The level spacing 
distribution is 
consistent with the Wigner surmise for 
the 
unitary metal, 
independent of the disorder strength. 
(Results are quantitatively the same as in Fig.~\ref{DiracLS}).
In addition, the 
multifractal
spectra show rather universal behavior. 
These 
are presented 
for various disorder strengths 
in Fig. \ref{FEMFS}. The singularity spectrum $f(\alpha)$ shows saturation for $\Delta_A\ge 0.8\pi$. 
The saturated spectrum is close to
\begin{align}\label{fpara}
	f(\alpha) = 2 - \frac{1}{4 \theta}(\alpha - 2 - \theta)^2,
\end{align}	
with $\theta\approx 0.26$.
This is the Legendre transform of 
the pure parabolic $\tau(q)$ spectrum in Eq.~(\ref{para}),   
which 
describes to a good approximation 
the multifractal spectrum for the integer quantum Hall plateau transition.\cite{Evers01,Evers08,Obuse08} 

\begin{figure}[ht]
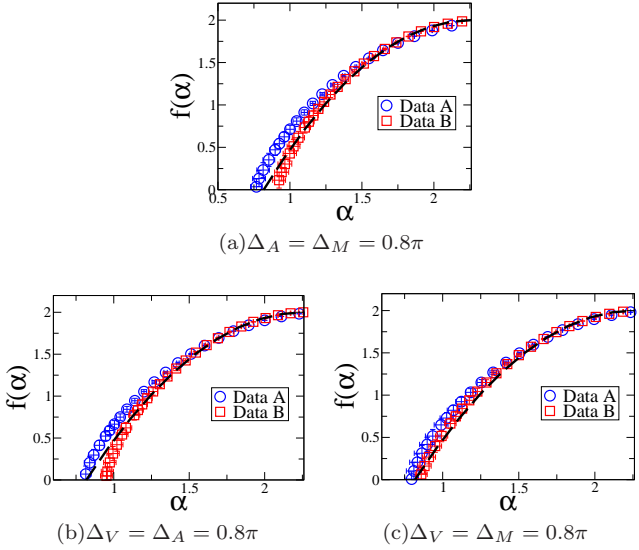

 \centering
 \subfigure[$\Delta_A=\Delta_M=0.8\pi$]{
\includegraphics[width=0.225\textwidth]{fa_A_M_08.eps}
   }\\[3mm]
 \subfigure[$\Delta_V=\Delta_A=0.8\pi$]{
\includegraphics[width=0.225\textwidth]{fa_V_A_08.eps}
   }
 \subfigure[$\Delta_V=\Delta_M=0.8\pi$]{
\includegraphics[width=0.225\textwidth]{fa_V_M_08.eps}
   }
\caption{The $f(\alpha)$ spectra of 
low
energy states 
for 
a 
single-valley
Dirac fermion with two kinds of disorder in MFD; 
$\mathcal{N}=40$. We perform 
an average over 80 realizations of the disorder.
$\Delta_V$ and $\Delta_M$ correspond to the disorder 
variance of scalar and mass potentials, respectively. 
$\xi=0.25r$ for all the cases, 
where $r=L/\mathcal{N}$. 
Data A is extracted from 
binning sizes 
$b=1$ and $b=2$. 
Data B is extracted from $b=2$ and $b=4$. 
The dashed line is the same as stated in Fig.~\ref{FEMFS}.}
\label{QHEMFS}
\end{figure}

Our result is the first numerical evidence for the delocalization of the finite energy 
states in the single-valley model based on the universal multifractal spectrum for the 
integer quantum Hall plateau transition.
For comparison, we also show $f(\alpha)$ for a single-valley Dirac fermion in the presence 
of two different types of disorder
in Fig.~\ref{QHEMFS}.

%%%%%%%%%%%%%%%%%%%%%%%%%%%%%%%%%%%%%%%%%%%%%%%%%%%%%%%%%%%%%%%%%%%%%%%%%%%%%%%%%%%%%%%%%
%%%%%%%%%%%%%%%%%%%%%%%%%%%%%%%%%%%%%%%%%%%%%%%%%%%%%%%%%%%%%%%%%%%%%%%%%%%%%%%%%%%%%%%%%
%%%%%%%%%%%%%%%%%%%%%%%%%%%%%%%%%%%%%%%%%%%%%%%%%%%%%%%%%%%%%%%%%%%%%%%%%%%%%%%%%%%%%%%%%
%%%%%%%%%%%%%%%%%%%%%%%%%%%%%%%%%%%%%%%%%%%%%%%%%%%%%%%%%%%%%%%%%%%%%%%%%%%%%%%%%%%%%%%%%
%%%%%%%%%%%%%%%%%%%%%%%%%%%%%%%%%%%%%%%%%%%%%%%%%%%%%%%%%%%%%%%%%%%%%%%%%%%%%%%%%%%%%%%%%
%%%%%%%%%%%%%%%%%%%%%%%%%%%%%%%%%%%%%%%%%%%%%%%%%%%%%%%%%%%%%%%%%%%%%%%%%%%%%%%%%%%%%%%%%
%%%%%%%%%%%%%%%%%%%%%%%%%%%%%%%%%%%%%%%%%%%%%%%%%%%%%%%%%%%%%%%%%%%%%%%%%%%%%%%%%%%%%%%%%
%%%%%%%%%%%%%%%%%%%%%%%%%%%%%%%%%%%%%%%%%%%%%%%%%%%%%%%%%%%%%%%%%%%%%%%%%%%%%%%%%%%%%%%%%
%%%%%%%%%%%%%%%%%%%%%%%%%%%%%%%%%%%%%%%%%%%%%%%%%%%%%%%%%%%%%%%%%%%%%%%%%%%%%%%%%%%%%%%%%
%%%%%%%%%%%%%%%%%%%%%%%%%%%%%%%%%%%%%%%%%%%%%%%%%%%%%%%%%%%%%%%%%%%%%%%%%%%%%%%%%%%%%%%%%
%%%%%%%%%%%%%%%%%%%%%%%%%%%%%%%%%%%%%%%%%%%%%%%%%%%%%%%%%%%%%%%%%%%%%%%%%%%%%%%%%%%%%%%%%
%%%%%%%%%%%%%%%%%%%%%%%%%%%%%%%%%%%%%%%%%%%%%%%%%%%%%%%%%%%%%%%%%%%%%%%%%%%%%%%%%%%%%%%%%
%%%%%%%%%%%%%%%%%%%%%%%%%%%%%%%%%%%%%%%%%%%%%%%%%%%%%%%%%%%%%%%%%%%%%%%%%%%%%%%%%%%%%%%%%
%%%%%%%%%%%%%%%%%%%%%%%%%%%%%%%%%%%%%%%%%%%%%%%%%%%%%%%%%%%%%%%%%%%%%%%%%%%%%%%%%%%%%%%%%
%%%%%%%%%%%%%%%%%%%%%%%%%%%%%%%%%%%%%%%%%%%%%%%%%%%%%%%%%%%%%%%%%%%%%%%%%%%%%%%%%%%%%%%%%
%%%%%%%%%%%%%%%%%%%%%%%%%%%%%%%%%%%%%%%%%%%%%%%%%%%%%%%%%%%%%%%%%%%%%%%%%%%%%%%%%%%%%%%%%

\section{Non-Abelian Vector Potential Dirac Fermions \label{Sec: NA}}

Bulk 
topological 
superconductors	
in classes CI and AIII 
can host multiple 
surface Dirac bands.
The number of species (or: ``valleys'') of Dirac fermions at the surface
is equal to the modulus of the corresponding bulk winding number $|\nu|$.\cite{Schnyder08}
For a superconductor with $|\nu| > 1$,
spin SU(2) and time-reversal invariant disorder manifests as 
a non-abelian valley vector potential in the low-energy surface
Dirac theory, 
which can mediate both intra- and intervalley scattering. 
This encodes the effects of charged impurities, vacancies, 
as well as corner and edge potentials on the surface.\cite{Foster14}

We focus on the two-valley model as the simplest example of Dirac fermions subject to 
non-abelian vector potentials.
The two-valley Dirac Hamiltonian 
is
\begin{align}\label{HAIII2}
	\mathcal{H}
	&=
	\int d^2\vex{x}\,
	\psi^{\dagger}(\vex{x})
	\left[-i\bm{\sigma}\cdot\Nabla+\bm\sigma\cdot\vex{A}_0(\vex{x})\right]
	\psi(\vex{x})\\
	&+
	\int d^2\vex{x}\,
	\psi^{\dagger}(\vex{x})
	\left[\sum_{a=x,y,z}\kappa_a\bm\sigma\cdot\vex{A}_a(\vex{x})\right]
	\psi(\vex{x}),
\end{align}
where $\vex{A}_{a}$ 
couples to the
valley 
space Pauli
matrix $\kappa_a$ 
($a \in \{x,y,z\}$), 
and $\vex{A}_0$ 
is an 
abelian vector potential,
as appears 
in the single valley case. 
We implement the random abelian and non-abelian
vector potentials
in the momentum space Dirac fermion (MFD) scheme described 
in Sec.~\ref{Sec: MFD}. 
The disorder variance for 
the 
non-abelian 
potential 
is 
denoted by $\Delta_{N}$. 
In the absence of the abelian vector potential, the system belongs to class 
 CI,\cite{Schnyder08,Schnyder09,Foster12} and can be realized at the 
surface of a spin SU(2) invariant topological superconductor.
A non-zero abelian potential couples to the U(1) spin current,
associated with the conserved component of spin.
[This is the U(1) charge of the Dirac quasiparticle field $\psi$, which carries
well-defined angular momentum but not electric charge.\cite{Foster14}]
When both the abelian and non-abelian vector potentials are present, 
the model resides in class AIII as in the single valley case. 
A topological superconductor in class AIII can be realized if time-reversal
and a remnant U(1) of the spin SU(2) symmetry is preserved in every realization of the
disorder, as might arise, e.g., through spin-triplet p-wave pairing.\cite{Foster08,Schnyder08} 

The problem of 2D Dirac fermions coupled to random vector potentials is 
exactly solvable by methods of conformal field theory;\cite{Nersesyan94,Nersesyan95,Mudry96,Caux96,Foster12}
for a review, see e.g.\ Ref.~\onlinecite{Foster14}. 
The relevant theory for a topological superconductor surface state with winding number $|\nu|$
is a Wess-Zumino-Witten model at level $|\nu|/2$ ($|\nu|$) in class CI (AIII).\cite{Schnyder09,Foster12,Foster14} 

For the system at the Wess-Zumino-Witten fixed point, the 
critical behavior of the global DoS\cite{Nersesyan94,Nersesyan95,Foster12} 
and the multifractal 
spectrum\cite{Mudry96,Caux96} 
of local density of states fluctuations   
can be calculated exactly.
For the
two-valley case, the dynamic critical exponent
is given by 
\begin{align}
	z
	=
	\frac{7}{4}+\frac{\Delta_A}{\pi}.
	\label{NA_dyn}
\end{align}
This result is independent of the non-abelian disorder
strength, and becomes universal when $\Delta_A \rightarrow 0$.
As in the abelian model, 
a freezing transition is expected to take place when 
$\Delta_A$ 
is larger than a certain threshold value (equal to $7 \pi /4$ for two valleys). 
The 
multifractal spectrum is exactly parabolic, up to termination. 
For two valleys, the parameter $\theta$ in Eqs.~(\ref{para}) and (\ref{fpara})
takes the value\cite{Mudry96,Caux96}
\begin{align}
	\theta
	=
	\frac{1}{4}+\frac{\Delta_A}{\pi}.
	\label{fa_NA}
\end{align}

\begin{figure}
\includegraphics[width=0.4\textwidth]{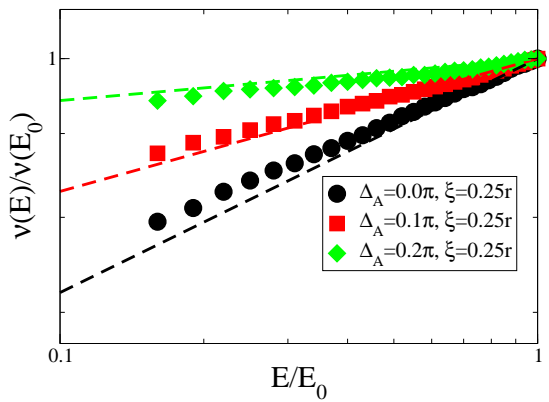}
\caption{The DoS near zero energy for 
2-valley class 
CI and AIII 
Dirac models, 
with $\mathcal{N}=40$. Dots are the numerical results from 400 energy 
levels, averaged over 
40 disorder 
realizations. 
The strength of the non-abelian SU(2) vector potential disorder 
$\Delta_N$ is fixed to $0.8\pi$ for all three cases; $r=L/\mathcal{N}$. 
The case(s) with $\Delta_A = 0$ ($\Delta_A > 0$) correspond
to class CI (AIII). 
The solid lines are the analytical 
result implied by Eqs.~(\ref{dos}) and (\ref{NA_dyn}).  
The data are rescaled so that the rightmost points are placed at 
the same position.}
\label{DoSAIIICI}
\end{figure}

\begin{figure}
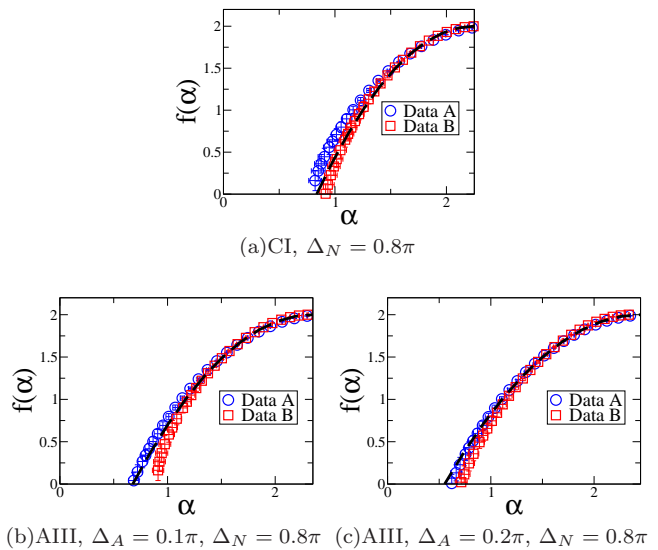

 \centering
 \subfigure[CI, $\Delta_{N}=0.8\pi$]{
\includegraphics[width=0.225\textwidth]{CI_fa_N40_g08_dis40.eps}
   }\\[3mm]
 \subfigure[AIII, $\Delta_A=0.1\pi$, $\Delta_{N}=0.8\pi$]{
\includegraphics[width=0.225\textwidth]{AIII_N40_ga01_g08_dis40.eps}
   }
 \subfigure[AIII, $\Delta_A=0.2\pi$, $\Delta_{N}=0.8\pi$]{
\includegraphics[width=0.225\textwidth]{AIII_N40_ga02_g08_dis40.eps}
   }
 \caption{The $f(\alpha)$ spectra of Dirac fermions with non-Abelian SU(2) 
vector potential in MFD. 
Here 
$\mathcal{N}=40$, $\xi=0.25r$ ($r=L/\mathcal{N}$), and 
we average over 40 disorder 
realizations for all three cases. The data are extracted from the numerical 
derivatives 
of 
the 
IPR. Data A is extracted from 
 binning sizes 
$b=1$ and $b=2$. Data B is extracted from $b=2$ and $b=4$. 
The solid lines are the analytical prediction
in 
Eqs.~(\ref{fpara}) and (\ref{fa_NA}). 
}
\label{faCIAII}
\end{figure}

We use MFD to 
compute $z$ and the multifractal spectrum
for two-valley surface states,
using grid sizes 
$\mathcal{N}=32$ to $48$. 
In Fig. \ref{DoSAIIICI}, the 
critical behavior of the DoS [related to $z$ via Eq.~(\ref{dos})] 
found numerically agrees well with the analytical prediction implied by 
Eq.~(\ref{NA_dyn}).
Moreover, the 
$f(\alpha)$ spectra 
shown in 
Fig.~\ref{faCIAII} are 
also close to the  
analytical predictions.

The numerical data shows good agreement with the conformal field theory
results. This appears to imply that the topology protects both the delocalization of the 
wavefunctions and the \emph{strict conformal invariance} of the surface. To understand this,
we consider a perturbation of the class CI and AIII Wess-Zumino-Witten models. 
In the conformal limit, the coefficient $1/\lambda$ of the gradient term in the non-abelian
bosonization of these theories is equal to the level $k$.\cite{CFTBook,Foster14}
If we deform $\lambda$ away from this, we get a non-conformal theory
(principle chiral model with a Wess-Zumino-Witten term).  
In the large-$k$ limit, the lowest order RG equations are given by\cite{Witten84,Guruswamy2000,Foster14}
\begin{subequations}
\begin{align}
	&\text{CI:} &\frac{d\lambda}{d l}=\lambda^2\left[1-(k\lambda)^2\right],\label{NA_RG_CI}\\
	&\text{AIII:} &\frac{d\lambda}{d l}=0,\,\, 
	\frac{d \Delta_A}{d l}= \pi \lambda^2\left[1-(k\lambda)^2\right].\label{NA_RG_AIII}
\end{align}
\end{subequations}

In class CI, the deformation is irrelevant: Eq.~(\ref{NA_RG_CI}) implies that the system flows 
back to the conformal limit ($\lambda = 1/k$). 
On the other hand, in class AIII Eq.~(\ref{NA_RG_AIII}) implies that the \emph{abelian disorder
variance} $\Delta_A$ becomes scale dependent whenever $\lambda \neq 1/k$. 
Although Eq.~(\ref{NA_RG_AIII}) can be obtained by perturbation theory in $\lambda \sim 1/k$,
valid in the limit $k \gg 1$, these results turn out to be exact.\cite{Guruswamy2000}
We conclude that any deformation away from the conformal limit in class AIII induces
a runaway flow of $\Delta_A$. As a result, one finds Gade-Wegner physics,\cite{Gade1991213,Gade1993499}
wherein the DoS assumes the strongly divergent form in Eq.~(\ref{DoS_Gade}).
The low-energy wavefunctions should always exhibit frozen multifractal spectra.

Although we are limited to small system sizes, we do not observe any signatures
of the Gade-Wegner scaling in class AIII. For example, the low-energy DoS vanishes 
for $0 < \Delta_A < \pi/4$, 
as indicated in Fig.~\ref{DoSAIIICI}. 
Moreover, the multifractal spectra in Fig.~\ref{faCIAII} are consistent with the parabolic spectra 
implied
by Eq.~(\ref{fa_NA}).
These suggest that the disordered Dirac theory in Eq.~(\ref{HAIII2})
flows under the renormalization group
directly to the AIII conformal field theory, without inducing the perturbation
$\lambda \neq 1/k$. 
As discussed in Refs.~\onlinecite{Foster14,Xie14}, this is consistent with the 
result\cite{Ludwig94,Tsvelik95,Ostrovsky06} that the Landauer (spin) conductance is universal for
non-interacting 2D Dirac fermions coupled to random vector potentials. It is then natural to 
interpret the coupling strength $\lambda=1/k$ of the Wess-Zumino-Witten theory as 
the inverse Landauer spin conductance. 
As discussed elsewhere,\cite{Xie14,Foster14} the lowest-order interaction
corrections to the conductance also vanish. These results suggest
the possibility that the surface state spin conductance of a topological
superconductor is truly universal (i.e., independent of both disorder and interactions), 
and provides a way to measure the bulk winding number directly via transport,
without modifying the surface\cite{Hasan10_RMP,Qi11_RMP} in some special way.

%%%%%%%%%%%%%%%%%%%%%%%%%%%%%%%%%%%%%%%%%%%%%%%%%%%%%%%%%%%%%%%%%%%%%%%%%%%%%%%%%%%%%%%%%
%%%%%%%%%%%%%%%%%%%%%%%%%%%%%%%%%%%%%%%%%%%%%%%%%%%%%%%%%%%%%%%%%%%%%%%%%%%%%%%%%%%%%%%%%
%%%%%%%%%%%%%%%%%%%%%%%%%%%%%%%%%%%%%%%%%%%%%%%%%%%%%%%%%%%%%%%%%%%%%%%%%%%%%%%%%%%%%%%%%
%%%%%%%%%%%%%%%%%%%%%%%%%%%%%%%%%%%%%%%%%%%%%%%%%%%%%%%%%%%%%%%%%%%%%%%%%%%%%%%%%%%%%%%%%
%%%%%%%%%%%%%%%%%%%%%%%%%%%%%%%%%%%%%%%%%%%%%%%%%%%%%%%%%%%%%%%%%%%%%%%%%%%%%%%%%%%%%%%%%
%%%%%%%%%%%%%%%%%%%%%%%%%%%%%%%%%%%%%%%%%%%%%%%%%%%%%%%%%%%%%%%%%%%%%%%%%%%%%%%%%%%%%%%%%
%%%%%%%%%%%%%%%%%%%%%%%%%%%%%%%%%%%%%%%%%%%%%%%%%%%%%%%%%%%%%%%%%%%%%%%%%%%%%%%%%%%%%%%%%
%%%%%%%%%%%%%%%%%%%%%%%%%%%%%%%%%%%%%%%%%%%%%%%%%%%%%%%%%%%%%%%%%%%%%%%%%%%%%%%%%%%%%%%%%
%%%%%%%%%%%%%%%%%%%%%%%%%%%%%%%%%%%%%%%%%%%%%%%%%%%%%%%%%%%%%%%%%%%%%%%%%%%%%%%%%%%%%%%%%
%%%%%%%%%%%%%%%%%%%%%%%%%%%%%%%%%%%%%%%%%%%%%%%%%%%%%%%%%%%%%%%%%%%%%%%%%%%%%%%%%%%%%%%%%
%%%%%%%%%%%%%%%%%%%%%%%%%%%%%%%%%%%%%%%%%%%%%%%%%%%%%%%%%%%%%%%%%%%%%%%%%%%%%%%%%%%%%%%%%
%%%%%%%%%%%%%%%%%%%%%%%%%%%%%%%%%%%%%%%%%%%%%%%%%%%%%%%%%%%%%%%%%%%%%%%%%%%%%%%%%%%%%%%%%
%%%%%%%%%%%%%%%%%%%%%%%%%%%%%%%%%%%%%%%%%%%%%%%%%%%%%%%%%%%%%%%%%%%%%%%%%%%%%%%%%%%%%%%%%
%%%%%%%%%%%%%%%%%%%%%%%%%%%%%%%%%%%%%%%%%%%%%%%%%%%%%%%%%%%%%%%%%%%%%%%%%%%%%%%%%%%%%%%%%
%%%%%%%%%%%%%%%%%%%%%%%%%%%%%%%%%%%%%%%%%%%%%%%%%%%%%%%%%%%%%%%%%%%%%%%%%%%%%%%%%%%%%%%%%
%%%%%%%%%%%%%%%%%%%%%%%%%%%%%%%%%%%%%%%%%%%%%%%%%%%%%%%%%%%%%%%%%%%%%%%%%%%%%%%%%%%%%%%%%

\section{Discussion \label{Sec: Disc}}

In this paper we have studied 
random vector potential Dirac fermions 
in 2D 
with one 
and two valleys. 
Both cases 
can be realized as the surface states of bulk 
topological superconductors.

For the single-valley model, we 
computed 
various physical properties for states in the 
low-energy 
chiral region, 
below and above the freezing transition
(i.e., for weak and strong disorder). 
Neither 
the level statistics 
nor the 
two-wavefunction correlations 
show a qualitative change at the freezing transition.
At strong disorder, level statistics remain approximately
Wigner-Dyson, and the overlap of the probability distributions for
different wavefunctions retains a power-law correlation in energy. 
The results 
imply 
that 
even the ``quasilocalized,'' highly 
rarefied wavefunctions 
in the strong disorder, frozen regime
are correlated in energy and obey generalized Chalker scaling. 
We want to emphasize that these critically delocalized wavefunctions 
are not the same as those near the mobility edge.\cite{Shklovskii93,Kravtsov94}
The crucial difference is that in the single valley model, \emph{all} states are delocalized
within the low-energy disordered Dirac region, even for strong disorder.

In addition to the low-energy physics of the single-valley model, we also 
investigated 
the states away from chiral region. We 
confirmed that the states at finite energies are delocalized 
based on their universal multifractal behavior. The multifractal 
spectrum of these states is well-approximated by that of 
the integer quantum Hall plateau transition. 
To our knowledge, this is the first numerical evidence to show the 
connection between finite-energy states and the plateau transition.

For 
the 
two-valley model, we 
demonstrated 
that 
Gade-Wegner 
scaling 	
does not 
occur 
for the AIII class. 
Our numerical results for the global DoS
and the multifractal spectra match well the 
predictions of 
conformal field theory.

We 
discussed 
two numerical methods, 
the momentum space Dirac formalism (MFD) and 
the 
 MDH lattice
model. MFD is a way to directly simulate 
the
Dirac fermion problem. It is useful to probe 
systems with a
vanishing DoS and weak multifractality. One can study both 
low-energy states and states away from the chiral region.
MFD is also suitable for simulating multiple valleys and random potentials.
The disadvantage is 
the restriction to 
relatively small system sizes.

The MDH\cite{Motrunich02} lattice model
is designed for studying single-valley Dirac fermions subject to a static 
random vector potential. The low-energy theory is 
described by 
Eq.~(\ref{lowE}), 
with
parametrically small mass terms
$m_{x,y}$.
The 
low-energy
properties 
including 
generalized Chalker scaling, 
the critical behavior of the DoS, and 
multifractal spectra are consistent with the analytical predictions 
for the single valley model 
over 
a 
substantial 
range of $\Delta_A$,
including the strong disorder regime above the freezing transition. 
On the other hand, 
the states far away from the chiral region are Anderson localized. 
The 
MDH 
lattice 
model 
might be realizable in 
artificial materials 
such as 
molecular graphene.\cite{Gomes12} 
The 
global 
DoS and multifractal spectrum 
of local DoS fluctuations
are both experimentally measurable quantities.

We close 
with open questions 
and 
future directions. The surface states for class DIII 
topological superconductors can also be described by 
real 
random vector potential Dirac 
(Majorana) fermions.\cite{Schnyder08,Foster14,Xie14} 
As in class AIII, it is 
important to understand whether conformal invariance is preserved 
for class DIII with three 
or 
more valleys.\cite{Foster14}
The non-abelian vector potential Dirac fermion 
in class CI shows universal behavior in the
DoS and 
multifractal spectrum. 
Constructing 
a non-abelian version of the MDH model on a lattice 
might allow the simulation of 
Dirac fermions with non-abelian vector potentials 
in 
artificial materials,
and would also allow efficient numerics for
much larger system sizes than we could access here using
the MFD approach. 

We have focused on typical multifractal spectra, obtained by disorder averaging 
the log of the inverse participation ratio (IPR). The freezing phenomena is related to 
rare extrema of a typical wavefunction. One can alternatively disorder average the IPR, 
and then take the log. This gives information about rare configurations of the disorder.
It will also be interesting to calculate the disorder-averaged IPR in order 
to verify the pre-freezing phenomenon.\cite{fyodorov09}

\section{Acknowledgments}

We thank Hong-Yi Xie for discussions on Chalker scaling. This work was supported by 
the Welch Foundation under Grant No.~C-1809.

%%%%%%%%%%%%%%%%%%%%%%%%%%%%%%%%%%%%%%%%%%%%%%%%%%%%%%%%%%%%%%%%%%%%%%%%%%%%%%%%
%%%%%%%%%%%%%%%%%%%%%%%%%%%%%%%%%%%%%%%%%%%%%%%%%%%%%%%%%%%%%%%%%%%%%%%%%%%%%%%%
%%%%%%%%%%%%%%%%%%%%%%%%%%%%%%%%%%%%%%%%%%%%%%%%%%%%%%%%%%%%%%%%%%%%%%%%%%%%%%%%
%%%%%%%%%%%%%%%%%%%%%%%%%%%%%%%%%%%%%%%%%%%%%%%%%%%%%%%%%%%%%%%%%%%%%%%%%%%%%%%%
%%%%%%%%%%%%%%%%%%%%%%%%%%%%%%%%%%%%%%%%%%%%%%%%%%%%%%%%%%%%%%%%%%%%%%%%%%%%%%%%
%%%%%%%%%%%%%%%%%%%%%%%%%%%%%%%%%%%%%%%%%%%%%%%%%%%%%%%%%%%%%%%%%%%%%%%%%%%%%%%%
%%%%%%%%%%%%%%%%%%%%%%%%%%%%%%%%%%%%%%%%%%%%%%%%%%%%%%%%%%%%%%%%%%%%%%%%%%%%%%%%
%%%%%%%%%%%%%%%%%%%%%%%%%%%%%%%%%%%%%%%%%%%%%%%%%%%%%%%%%%%%%%%%%%%%%%%%%%%%%%%%
%%%%%%%%%%%%%%%%%%%%%%%%%%%%%%%%%%%%%%%%%%%%%%%%%%%%%%%%%%%%%%%%%%%%%%%%%%%%%%%%
%%%%%%%%%%%%%%%%%%%%%%%%%%%%%%%%%%%%%%%%%%%%%%%%%%%%%%%%%%%%%%%%%%%%%%%%%%%%%%%%
%%%%%%%%%%%%%%%%%%%%%%%%%%%%%%%%%%%%%%%%%%%%%%%%%%%%%%%%%%%%%%%%%%%%%%%%%%%%%%%%
%%%%%%%%%%%%%%%%%%%%%%%%%%%%%%%%%%%%%%%%%%%%%%%%%%%%%%%%%%%%%%%%%%%%%%%%%%%%%%%%

\appendix

%%%%%%%%%%%%%%%%%%%%%%%%%%%%%%%%%%%%%%%%%%%%%%%%%%%%%%%%%%%%%%%%%%%%%%%%%%%%%%%%
%%%%%%%%%%%%%%%%%%%%%%%%%%%%%%%%%%%%%%%%%%%%%%%%%%%%%%%%%%%%%%%%%%%%%%%%%%%%%%%%
%%%%%%%%%%%%%%%%%%%%%%%%%%%%%%%%%%%%%%%%%%%%%%%%%%%%%%%%%%%%%%%%%%%%%%%%%%%%%%%%
%%%%%%%%%%%%%%%%%%%%%%%%%%%%%%%%%%%%%%%%%%%%%%%%%%%%%%%%%%%%%%%%%%%%%%%%%%%%%%%%
%%%%%%%%%%%%%%%%%%%%%%%%%%%%%%%%%%%%%%%%%%%%%%%%%%%%%%%%%%%%%%%%%%%%%%%%%%%%%%%%
%%%%%%%%%%%%%%%%%%%%%%%%%%%%%%%%%%%%%%%%%%%%%%%%%%%%%%%%%%%%%%%%%%%%%%%%%%%%%%%%
%%%%%%%%%%%%%%%%%%%%%%%%%%%%%%%%%%%%%%%%%%%%%%%%%%%%%%%%%%%%%%%%%%%%%%%%%%%%%%%%
%%%%%%%%%%%%%%%%%%%%%%%%%%%%%%%%%%%%%%%%%%%%%%%%%%%%%%%%%%%%%%%%%%%%%%%%%%%%%%%%
%%%%%%%%%%%%%%%%%%%%%%%%%%%%%%%%%%%%%%%%%%%%%%%%%%%%%%%%%%%%%%%%%%%%%%%%%%%%%%%%
%%%%%%%%%%%%%%%%%%%%%%%%%%%%%%%%%%%%%%%%%%%%%%%%%%%%%%%%%%%%%%%%%%%%%%%%%%%%%%%%
%%%%%%%%%%%%%%%%%%%%%%%%%%%%%%%%%%%%%%%%%%%%%%%%%%%%%%%%%%%%%%%%%%%%%%%%%%%%%%%%
%%%%%%%%%%%%%%%%%%%%%%%%%%%%%%%%%%%%%%%%%%%%%%%%%%%%%%%%%%%%%%%%%%%%%%%%%%%%%%%%

\section{Low Energy Theory of Real Random Hopping $\pi$-Flux Model \label{App: Pi-flux}}

We discuss how to derive Dirac fermions in the real random hopping $\pi$-flux model in this 
appendix.
The lattice model belongs to the class BDI in the Altland-Zirnbauer 
classification.\cite{Evers08_RMP} 

We first consider the real random hopping $\pi$-flux model. The Hamiltonian is given by
\begin{align}
	\nonumber
	\mathcal{H}=\sum_{\vex r}\Big[&t_{\vex{r},\vex{r}+\hat{x}}(-1)^{\vex r\cdot\hat{y}}c^{\dagger}(\vex r)c(\vex r+\hat{x})\\
	+&t_{\vex{r},\vex{r}+\hat{y}}c^{\dagger}(\vex r)c(\vex r+\hat{y})+\text{h.c}\Big],\label{H_pi}
\end{align}
The $\pi$-flux lattice contains two 
sites per unit cell, 
$\alpha$ and $\beta$. 
These cannot be chosen in the same way as the sublattice labels $A$ and $B$ 
shown in Fig. \ref{MDH_lattice}. 
The primitive vectors are $\vex{t}_1=\hat{x}$ and $\vex{t}_2=2\hat{y}$. The lattice constant is set to unity. 
We 
label the sites via 
$\vex{r}_{\alpha}=(n,2m)$, $\vex{r}_{\beta}=(n,2m+1)$, 
with $n,m \in \mathbb{Z}$,
and we define 
$a(\vex{r}_{\alpha})\equiv c(\vex{r}_{\alpha})$, 
and $b(\vex{r}_{\beta})\equiv c(\vex{r}_{\beta})$. 

In the clean limit ($t_{\vex{r},\vex{r}+\hat{x}}=t_{\vex{r},\vex{r}+\hat{y}}=t$), 
the Hamiltonian in 
momentum 
space is
\begin{align*}
	\mathcal{H}=2t\int_{\vex{k}\in \text{B.Z.}}
	\Phi^{\dagger}(\vex{k})\left[\begin{array}{cc}
	-\cos(k_x) & \cos(k_y)\\
	\cos(k_y) & \cos(k_x)
	\end{array}\right]\Phi(\vex{k}),
\end{align*}
where $\Phi^{\dagger}(\vex{k})=\left[a^{\dagger}(\vex{k}),\,\, b^{\dagger}(\vex{k})\right]$. 
The dispersion is 
\[
	\omega_{\vex{k}}=\pm 2t\sqrt{\cos^2(k_x)+\cos^2(k_y)}.
\]

The clean $\pi$-flux model can be described by two valleys of decoupled massless Dirac fermions. 
The distinct Dirac points
are 
$\vex{K}_+=(\frac{\pi}{2},\frac{\pi}{2})$ and $\vex{K}_-=(-\frac{\pi}{2},\frac{\pi}{2})$. 
The reciprocal vectors of the lattice problem are $\vex{Q}_1=2\pi\hat{x}$ and $\vex{Q}_1=\pi\hat{y}$.\\

In the low energy limit, only degrees of freedom near 
the 
Dirac points play important roles. We therefore use 
the 
valley decomposition of 
the 
fields,
\begin{align}
	\nonumber 
	a(\vex{r}_{\alpha})
	\approx 
	e^{i\vex{K}_+\cdot\bm{r}_{\alpha}}a_{+}(\vex{r}_{\alpha})
	+
	e^{i\vex{K}_-\cdot\vex{r}_{\alpha}}a_{-}(\vex{r}_{\alpha}),
	\\
	b(\vex{r}_{\beta})
	\approx 
	e^{i\vex{K}_+\cdot\vex{r}_{\beta}}b_{+}(\vex{r}_{\beta})
	+
	e^{i\vex{K}_-\cdot\vex{r}_{\beta}}b_{-}(\vex{r}_{\beta}),
	\label{c_LE}
\end{align}
where $+$ and $-$ subscripts specify the low energy degrees of freedom in the vicinity of Dirac points $\vex{K}_+$ and $\vex{K}_-$.

Fermion 
bilinears 
that appear 
in the $\pi$-flux model 
include 
$a^{\dagger}(\vex{r}_{\alpha})a(\vex{r}_{\alpha}\pm\hat{x})$, 
$b^{\dagger}(\vex{r}_{\beta})b(\vex{r}_{\beta}\pm\hat{x})$, 
$a^{\dagger}(\vex{r}_{\alpha})b(\vex{r}_{\alpha}\pm\hat{y})$, and 
$b^{\dagger}(\vex{r}_{\beta})a(\vex{r}_{\beta}\pm\hat{y})$. 
We perform 
the 
valley decomposition and Taylor expansion for all the bilinears. For example,
\begin{align*}
	a^{\dagger}(\vex{r}_{\alpha}) & a(\vex{r}_{\alpha}\pm\hat{x})
	+
	a^{\dagger}(\vex{r}_{\alpha}\pm\hat{x})a(\vex{r}_{\alpha})
	\\
	&\approx
	i\left[
	\begin{aligned}
	&\,
	a_+^{\dagger}(\partial_xa_+)-(\partial_xa_+^{\dagger})a_+
	\\
	&\,-
	a_-^{\dagger}(\partial_xa_-)+(\partial_xa_-^{\dagger})a_-
	\end{aligned}
	\right]_{\vex{r}_{\alpha}}
	\\
	&\mp 
	(2i)(-1)^{\vex{r}_{\alpha}\cdot\hat{x}}
	\left[
	a^{\dagger}_+a_-
	-
	a^{\dagger}_-a_+
	\right]_{\vex{r}_{\alpha}}.
\end{align*}
All the bilinear terms contain 
the 
staggered factor along 
the 
$x$ direction, $(-1)^{\vex{r}\cdot\hat{x}}$. 
It suggests that the minimum cell for constructing the low energy theory is a 
$2$-by-$2$ block. 

In the presence of disorder, the
hopping terms in Eq. (\ref{H_pi}) can be viewed as 
$t_{\vex r,\vex r'} = t + \delta t_{\vex r,\vex r'}$, where 
$\delta t_{\vex r,\vex r'}$ is a zero-mean random variable. In the clean limit, 
the 
low-energy 
Hamiltonian is
\begin{align*}
	\mathcal{H}_0
	= 
	2t\int_{\vex{x}}\psi^{\dagger}(\vex{x})\left[-i\sigma_z\kappa_z\partial_x+i\sigma_x\partial_y\right]\psi(\vex{x}),
\end{align*}
where
\begin{align*}
\psi=\left[\begin{array}{c}
a_{+}\\
b_{+}\\
a_{-}\\
b_{-}
\end{array}\right].
\end{align*}
Here the 
$\sigma$'s are 
Pauli matrices acting on 
($a/b$) space, and $\kappa$'s are the Pauli matrices on 
valley ($+/-$) space.
The disorder induces the appearance of vector potential and mass terms, 
\begin{align}
	\label{deltaH-pi}
	\delta\mathcal{H}
	\approx 
	2t
	\int_{\vex{x}}\psi^{\dagger}\left[A_x\sigma_y\kappa_x+A_y\kappa_y+m_x\sigma_y+m_y\sigma_z\kappa_y\right]\psi.
\end{align}
Note that the mass terms $m_x$ and $m_y$ commute 
with 
the vector potential
components, 
but anticommute 
with 
the kinetic term.

Now we are in the position to impose the correlated 
random 
hopping pattern of 
the 
MDH model.\cite{Motrunich02} 
The MDH pattern in 
the
$\pi$-flux model is listed below. 
In a 2-by-2 block associated with position $\vex{R}$,
the hopping elements in Eq.~(\ref{H_pi}) are assigned as 
\begin{align*}
	t_{\vex{R},\vex{R}\pm \hat{x}}&=e^{V(\vex{R})}te^{-V(\vex{R}\pm\hat{x})},\\[2mm]
	t_{\vex{R},\vex{R}\pm \hat{y}}&=e^{V(\vex{R})}te^{-V(\vex{R}\pm\hat{y})},\\[2mm]
	t_{\vex{R}+\hat{y},\vex{R}+\hat{y}\pm\hat{x}}&=e^{V(\vex{R}+\hat{y}\pm \hat{x})}te^{-V(\vex{R}+\hat{y})},\\[2mm]
	t_{\vex{R}+\hat{x},\vex{R}+\hat{x}\pm\hat{y}}&=e^{V(\vex{R}+\hat{x}\pm \vec{y})}te^{-V(\vex{R}+\hat{x})},
\end{align*}
where $\vex R=(2n,2m)$, $n$ and $m$ are integers. $V(\vex{y})$ is a random surface obeying Eq. (\ref{logCor}).
The low energy theory for 
the MDH model 
on $\pi$-flux lattice 
is given by 
\begin{align*}
	\mathcal{H}
	=
	\mathcal{H}_0
	+
	2t
	\int_{\vex{x}}
	\psi^{\dagger}
	\left[(\partial_yV)\sigma_y\kappa_x+(\partial_xV)\kappa_y\right]
	\psi.
\end{align*}
The mass terms in Eq.~(\ref{deltaH-pi}) vanish up to second order derivatives
in $V$, after we 
we coarse grain a 2-by-2 block in the lattice model at each 
position $\vex{R}$.

The derived low energy theory is nothing but Eq. (\ref{lowE}) after applying the following basis rotation,
\begin{align*}
	\psi\rightarrow\frac{1}{\sqrt{2}}\left(1+i\sigma_x\kappa_z\right)\frac{1}{\sqrt{2}}\left(1+i\kappa_y\right)\frac{1}{\sqrt{2}}\left(1+i\sigma_z\right)\psi.
\end{align*}

As a comparison, we also briefly discuss the MDH model on 
the 
honeycomb lattice. 
The hopping amplitudes can be generated via Eq.~(\ref{t_MDH}). 
The low energy theory 
reads
\begin{align*}
	\mathcal{H}\approx \frac{3t}{2}\int_{\vex{x}}\psi^{\dagger}\left[-i\sigma_x\partial_x-i\sigma_y\partial_y+\vex{A}\cdot\vex{\sigma}\mu_z\right]\psi,
\end{align*}
where the basis convention for 
the 
honeycomb lattice is
\begin{align}
\psi=\left[\begin{array}{c}
a_{+}\\
b_{+}\\
b_{-}\\
-a_{-}
\end{array}\right]\label{honeycomb_psi}
\end{align}
($a$ and $b$ label the triangular sublattices). 

The mass terms are related to the Kekul\'e patterns in the honeycomb lattice\cite{Hou07}.
A minimum 6-site hexagon is needed for performing the coarse graining procedure contrary to 
a 4-site square block for $\pi$-flux lattice. In this aspect, the MDH model on 
the 
honeycomb lattice will require 
a 
larger system in order to avoid deviations generated by non-zero masses. 
This is consistent with what we report for the numerical DoS in 
Fig.~\ref{DoSL256}, 
where 
results obtained for 
the
MDH $\pi$-flux and honeycomb lattices
are compared for equal system sizes.

%%%%%%%%%%%%%%%%%%%%%%%%%%%%%%%%%%%%%%%%%%%%%%%%%%%%%%%%%%%%%%%%%%%%%%%%%%%%%%%%
%%%%%%%%%%%%%%%%%%%%%%%%%%%%%%%%%%%%%%%%%%%%%%%%%%%%%%%%%%%%%%%%%%%%%%%%%%%%%%%%
%%%%%%%%%%%%%%%%%%%%%%%%%%%%%%%%%%%%%%%%%%%%%%%%%%%%%%%%%%%%%%%%%%%%%%%%%%%%%%%%
%%%%%%%%%%%%%%%%%%%%%%%%%%%%%%%%%%%%%%%%%%%%%%%%%%%%%%%%%%%%%%%%%%%%%%%%%%%%%%%%
%%%%%%%%%%%%%%%%%%%%%%%%%%%%%%%%%%%%%%%%%%%%%%%%%%%%%%%%%%%%%%%%%%%%%%%%%%%%%%%%
%%%%%%%%%%%%%%%%%%%%%%%%%%%%%%%%%%%%%%%%%%%%%%%%%%%%%%%%%%%%%%%%%%%%%%%%%%%%%%%%
%%%%%%%%%%%%%%%%%%%%%%%%%%%%%%%%%%%%%%%%%%%%%%%%%%%%%%%%%%%%%%%%%%%%%%%%%%%%%%%%
%%%%%%%%%%%%%%%%%%%%%%%%%%%%%%%%%%%%%%%%%%%%%%%%%%%%%%%%%%%%%%%%%%%%%%%%%%%%%%%%
%%%%%%%%%%%%%%%%%%%%%%%%%%%%%%%%%%%%%%%%%%%%%%%%%%%%%%%%%%%%%%%%%%%%%%%%%%%%%%%%
%%%%%%%%%%%%%%%%%%%%%%%%%%%%%%%%%%%%%%%%%%%%%%%%%%%%%%%%%%%%%%%%%%%%%%%%%%%%%%%%
%%%%%%%%%%%%%%%%%%%%%%%%%%%%%%%%%%%%%%%%%%%%%%%%%%%%%%%%%%%%%%%%%%%%%%%%%%%%%%%%
%%%%%%%%%%%%%%%%%%%%%%%%%%%%%%%%%%%%%%%%%%%%%%%%%%%%%%%%%%%%%%%%%%%%%%%%%%%%%%%%

\section{Random phase Disorder\label{App: CorrDis}}

In this appendix we discuss our parametrization of the disorder potentials employed in this paper. 
In particular, we 
show how to realize the correlated disorder with the random phase method
(discussed below). 
Consider a real-valued disorder potential, $B(\vex{x})$, satisfying
\begin{align}
	\langle B(\vex{x})\rangle_{\text{dis}}&=0,\label{zeromean}\\
	\langle B(\vex{x}+\vex{R})B(\vex{x})\rangle_{\text{dis}}&=\Delta_B \, \mathcal{K}(\vex{R})\label{var},
\end{align}
where $\langle\dots\rangle_{\text{dis}}$ denotes disorder average, 
$\Delta_B$ indicates the strength of the disorder potential, 
and $\mathcal{K}(\bm{R})$ is a normalized real-valued distribution in the position space. 

In the infinite size limit, one can exchange the disorder average $\langle\dots\rangle_{\text{dis}}$ 
by the spatial average $\langle\dots\rangle_{\bm{x}}$. 
In a finite system, 
we need to be careful about which 
scheme is employed. 
For 
Gaussian correlated disorder $G(\vex{x})$ in 2D,
one can 
parametrize the potential in terms of randomly positioned 
impurities with a Gaussian scattering profile,\cite{Bardarson07,Nomura07_DF} 
\begin{align*}
	G(\vex{x})
	=
	\frac{1}{2\pi s^2}
	\left[
	\sum_{j=1}^{N_+}
		e^{-\frac{\left(\vex{x}-\vex{y}^+_{j}\right)^2}{2s^2}}
	-
	\sum_{j=1}^{N_-}
	e^{-\frac{(\vex{x}-\vex{y}^-_{j})^2}{2s^2}}
	\right].
\end{align*}
In this equation, the 
$\vex{y}$'s indicate the positions of the impurities, 
and 
$N_+$ and $N_-$ are the numbers of positive charged and negative charged impurities. 
The disorder profile is determined by the configuration of 
the 
$\vex{y}$'s. For the zero mean
case, we choose $N_+=N_-=N$. 
The 
$G(\vex{x})$ generated 
in this way 
satisfies the following properties:
\begin{align*}
	\langle G(\vex{x})\rangle_{\{\vex{y}\}}
	&=0,
	\\
	\langle G(\vex{x}+\vex{R})G(\vex{x})\rangle_{\{\vex{y}\}}
	&=
	\frac{2N_+}{L^2}
	\left[\frac{1}{2\pi(\sqrt{2}s)^2} e^{-\frac{\vex{R}^2}{2(\sqrt{2}s)^2}}-\frac{1}{L^2}\right], 
\end{align*}
where 
\begin{align*}
	\langle f(\{\vex{y}\})\rangle_{\{\vex{y}\}}
	\equiv 
	\prod_{i}\left[\frac{1}{L^2}\int d^2\vex{y}_i\right]f(\{\vex{y}\}).
\end{align*}
The strength of the disorder potential is determined by the 
total 
density of the 
scatters $2N/L^2$. Moreover, there is a $\sqrt{2}$-enhancement in the resultant 
Gaussian correlation length. 

In a fixed disorder realization, the 
Fourier components of $G(\vex x)$ 
are given by
\begin{align*}
	\widetilde{G}_{\vex m\neq(0,0)}
	=
	e^{-\frac{1}{2}(\frac{2\pi}{L}\vex ms)^2}
	\left[
	\sum_{j=1}^{N}e^{i\frac{2\pi}{L}\vex{m}\cdot\vex{y^+_j}}
	-
	\sum_{j=1}^{N}e^{i\frac{2\pi}{L}\vex{m}\cdot\vex{y^-_j}}
	\right].
\end{align*}
When $N$ is sufficiently large, the 
term in 
square 
brackets 
can be approximated by a random phase term
\begin{align}
	\left[
	\sum_{j=1}^{N}e^{i\frac{2\pi}{L}\vex{m}\cdot\vex{y^+_j}}
	-
	\sum_{j=1}^{N}e^{i\frac{2\pi}{L}\vex{m}\cdot\vex{y^-_j}}
	\right]
	\approx
	\sqrt{2N}e^{i\phi_{\vex{m}}},\label{to_Ran_phase}
\end{align}
where $\phi_{-\vex{m}}=-\phi_{\vex{m}}$ for $\vex m\neq\vex 0$.

The scheme discussed above is limited to certain specific correlation profiles. 
For 
the
long-ranged correlated disorder potentials in Eq.~(\ref{logCor}), 
one needs to use a more general approach to generate randomness.

In the rest of the appendix, we focus on constructing disorder 
potentials by assigning random phases.
Instead of working 
with the 
conditions in 
Eqs.~(\ref{zeromean}) and (\ref{var}) 
directly, we replace the disorder average by the spatial average. Therefore, $B(\vex{x})$ satisfies
\begin{align}
\langle B(\vex{x})\rangle_{\vex{x}}&=0,\label{xzeromean}\\
\langle B(\vex{x}+\vex{R})B(\vex{x})\rangle_{\vex{x}}&=\Delta_B\mathcal{K}(\vex{R})\label{xvar},
\end{align}
where
\begin{align*}
	\langle f(\vex{x})\rangle_{\vex{x}}\equiv L^{-2}\int d^2{\vex{x}}\,f(\vex{x}).
\end{align*}

The zero-mean condition [Eq.~(\ref{xzeromean})] indicates that $\widetilde{B}_{\vex{n}=\vex 0}$ vanishes, where $\widetilde{B}_{\vex{n}}$ is the Fourier component of $B(\vex{x})$. The condition in Eq. (\ref{xvar}),
\begin{align*}
	&\langle B(\vex{x})B(\vex{x}+\bm{R})\rangle_{\vex{x}}=\Delta_B\mathcal{K}(\vex{R})\\[2mm]
	\rightarrow&\frac{1}{L^4} \sum_{\vex{m}}e^{i\Delta_k\vex{m}\cdot\vex{R}}\widetilde{B}_{-\vex{m}}\widetilde{B}_{\vex{m}}
	=\frac{\Delta_B}{L^2}\sum_{\vex{m}}e^{i\Delta_k\vex{m}\cdot\vex{R}}\widetilde{\mathcal{K}}_{\vex{m}}\\
	\rightarrow&\widetilde{B}_{-\vex{m}}\,\widetilde{B}_{\vex{m}}=\left|\widetilde{B}_{\vex{m}}\right|^2=L^2\Delta_B\widetilde{\mathcal{K}}_{\vex{m}}
\end{align*}
where we have used $\widetilde{B}_{\vex{m}}^*=\widetilde{B}_{-\vex{m}}$.

Assuming that $\widetilde{\mathcal{K}}_{\vex{m}}$ is real and non-negative, the disorder potential in the momentum space satisfies
\begin{align*}
	\widetilde{B}_{\vex m= 0}&=0,\\
	\widetilde{B}_{\vex{m}\neq 0}&=L\sqrt{\Delta_B}\left(\widetilde{\mathcal{K}}_{\vex{m}}\right)^{\frac{1}{2}}e^{i\theta_{\vex m}},
\end{align*}
where $\theta_{\vex m}$ is an uniform random variable from $0$ to $2\pi$ and $\theta_{-\vex m}=-\theta_{\vex m}$. 
The disorder average can be performed by averaging over $\theta$'s. The potential $B(\vex{x})$ constructed this way 
satisfies the following equations:
\begin{align*}
	\langle B(\vex{x})\rangle_{\{\theta\}}&=0,\\
	\langle B(\vex{x}+\vex{R})B(\vex{x})\rangle_{\{\theta\}}&=\Delta_B\left[\mathcal{K}(\vex{R})-\frac{1}{L^2}\right],
\end{align*}
where
\begin{align*}
	\langle f(\{\theta\})\rangle_{\{\theta\}}=\prod_{i}\left[\int_0^{2\pi}\frac{d\theta_i}{2\pi}\right]f(\{\theta\}).
\end{align*}
The $\prod_i$ in the above equation runs over all the independent $\theta_{i}$. The configuration 
of $\theta_{\vex m}$ characterizes the disordered potential. 
The finite size correction is similar to the 
random-position
impurity scheme discussed earlier.

The random phase method is particularly efficient in the MFD 
scheme
because the randomness is directly assigned to the Fourier mode, 
rather than the position space profile. 
This scheme also 
allows
us to simulate Eq.~(\ref{logCor}).


\begin{thebibliography}{67}
\bibitem{footnote--xival}
	The chosen value of the correlation length $\xi=0.25r$ relates the former
	to the fixed inverse ultraviolet cutoff $r$. The latter would be set relative
	to the lattice spacing in a microscopic model. 
	The value of $\xi$ should be small compared to the system size $L$, but finite because 
	the Gaussian correlation is a regularization of the white-noise condition in Eq.~(\ref{WN}).
	A slightly smaller or larger ratio than $0.25$ gives the same result; this is a
	stable parameter region.
\bibitem[{mas()}]{mass_term}
\bibinfo{note}{The mass terms in the low energy theory are scaled with
  $\sqrt{\Delta_A}L^{-2}$, up to logarithmic corrections in $L$.}
\end{thebibliography}
\end{document}